\documentclass[5p,authoryear]{elsarticle} 

\usepackage{framed,multirow}
\usepackage{amssymb}
\usepackage{latexsym}
\usepackage{xcolor}
\usepackage{lineno,hyperref}
\modulolinenumbers[5]
\usepackage{amsmath}
\usepackage{verbatim}
\usepackage{multirow}
\usepackage{color,graphicx}
\usepackage{arydshln}
\usepackage{hyperref}
\usepackage{url}
\usepackage{footnote}
\usepackage{bm}
\usepackage{cite}
\usepackage{subfigure}
\usepackage{verbatim}
\usepackage{ulem}
\newcommand{\tabincell}[2]{\begin{tabular}{@{}#1@{}}#2\end{tabular}}
\usepackage{arydshln}
\usepackage[figuresright]{rotating}

\providecommand{\leireftb}[1]{Table~\ref{#1}}
\providecommand{\leireffig}[1]{Figure~\ref{#1}}

\journal{Medical Image Analysis}

\begin{document}

\begin{frontmatter}
\title{MyoPS: A Benchmark of Myocardial Pathology Segmentation Combining Three-Sequence Cardiac Magnetic Resonance Images}

\author{Lei Li$^{1,2\dagger}$}  
\author{Fuping Wu$^{1\dagger}$}   
\author{Sihan Wang$^{1\dagger}$} 
\author{Xinzhe Luo$^{1}$} 
\author{Carlos Martín-Isla$^{3}$}
\author{Shuwei Zhai$^{5}$} 
\author{Jianpeng Zhang$^{6}$} 
\author{Yanfei Liu$^{7}$} 
\author{Zhen Zhang$^{9}$} 
\author{Markus J. Ankenbrand$^{10}$}
\author{Haochuan Jiang$^{11,12}$} 
\author{Xiaoran Zhang$^{13}$} 
\author{Linhong Wang$^{15}$} 
\author{Tewodros Weldebirhan Arega$^{16}$} 
\author{Elif Altunok$^{17}$} 
\author{Zhou Zhao$^{18}$} 
\author{Feiyan Li$^{15}$}
\author{Jun Ma$^{19}$} 
\author{Xiaoping Yang$^{20}$} 
\author{Elodie Puybareau$^{18}$} 
\author{Ilkay Oksuz$^{17}$} 
\author{Stephanie Bricq$^{16}$} 
\author{Weisheng Li$^{15}$}
\author{Kumaradevan Punithakumar$^{14}$} 
\author{Sotirios A. Tsaftaris$^{11}$} 
\author{Laura M. Schreiber$^{10}$}
\author{Mingjing Yang$^{9}$} 
\author{Guocai Liu$^{7,8}$} 
\author{Yong Xia$^{6}$} 
\author{Guotai Wang$^{5}$} 
\author{Sergio Escalera$^{3,4}$}
\author{Xiahai~Zhuang$^{1}$*} \ead[url]{Senior and corresponding author: http://www.sdspeople.fudan.edu.cn/zhuangxiahai/}

\address{$^{1}$School of Data Science, Fudan University, Shanghai, China\\[0.5ex]
$^{2}$School of Biomedical Engineering, Shanghai Jiao Tong University, Shanghai, China \\
$^{3}$Departament de Matemàtiques \& Informàtica, Universitat de Barcelona, Barcelona, Spain \\
$^{4}$Computer Vision Center, Universitat Autònoma de Barcelona, Spain \\
$^{5}$School of Mechanical and Electrical Engineering, University of Electronic Science and Technology of China, Chengdu, China \\
$^{6}$School of Computer Science and Engineering, Northwestern Polytechnical University, Xi'an, China \\
$^{7}$College of Electrical and Information Engineering, Hunan University, Changsha, China \\
$^{8}$National Engineering Laboratory for Robot Visual Perception and Control Technology, Changsha, China \\
$^{9}$College of Physics and Information Engineering, Fuzhou University, Fuzhou, China\\
$^{10}$Chair of Molecular and Cellular Imaging, Comprehensive Heart Failure Center, Wuerzburg University Hospitals, Wuerzburg, Germany \\
$^{11}$School of Engineering, University of Edinburgh, Edinburgh, UK \\
$^{12}$School of Robotics, Xi’an Jiaotong-Liverpool University, Suzhou, China \\
$^{13}$Department of Electrical and Computer Engineering, University of California, Los Angeles, USA \\
$^{14}$Department of Radiology and Diagnostic Imaging, University of Alberta, Edmonton, Canada \\
$^{15}$Chongqing Key Laboratory of Image Cognition, Chongqing University of Posts and Telecomm-unications, Chongqing, China\\
$^{16}$ImViA Laboratory, Université Bourgogne Franche-Comté, Dijon, France \\
$^{17}$Computer Engineering Department, Istanbul Technical University, Istanbul, Turkey \\
$^{18}$EPITA Research and Development Laboratory (LRDE), Le Kremlin-Bicêtre, France \\
$^{19}$Department of Mathematics, Nanjing University of Science and Technology, Nanjing, China \\
$^{20}$Department of Mathematics, Nanjing University, Nanjing, China 
}

\begin{abstract} 

Assessment of myocardial viability is essential in diagnosis and treatment management of patients suffering from myocardial infarction, and  classification of pathology on myocardium is the key to this assessment.
This work defines a new task of medical image analysis, i.e., to perform myocardial pathology segmentation (MyoPS) combining three-sequence cardiac magnetic resonance (CMR) images, which was first proposed in the MyoPS challenge, in conjunction with MICCAI 2020. 
The challenge provided 45 paired and pre-aligned CMR images, allowing algorithms to combine the complementary information from the three CMR sequences for pathology segmentation.
In this article, we provide details of the challenge, survey the works from fifteen participants and interpret their methods according to five aspects, i.e., preprocessing, data augmentation, learning strategy, model architecture and post-processing.
In addition, we  analyze the results with respect to different  factors, in order to examine the key obstacles and explore potential of solutions, as well as to provide a benchmark for future research.
We conclude that while promising results have been reported, the research is still in the early stage, and more in-depth exploration is needed before a successful application to the clinics. 
Note that MyoPS data and evaluation tool continue to be publicly available upon registration via its homepage {(www.sdspeople.fudan.edu.cn/zhuangxiahai/0/myops20/)}. 

\end{abstract}

\end{frontmatter}



\section{Introduction}

\begin{figure*}[t]\center
 \includegraphics[width=1.0\textwidth]{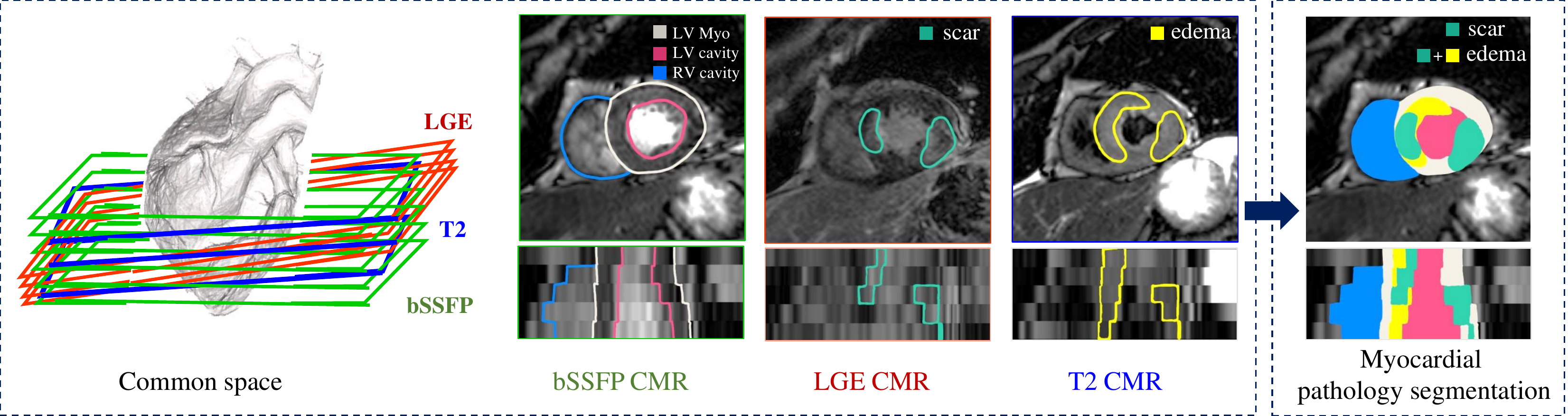}\\[-2ex]
   \caption{Visualization of myocardial pathology segmentation combining three-sequence cardiac magnetic resonance images acquired from the same patient (figure designed referring to \citet{journal/pami/Zhuang2019}).}
\label{fig:intro:MyoPS}\end{figure*}

\subsection{Clinical background}\label{intro:background}
Myocardial infarction (MI) is a major cause of mortality and disability worldwide \citep{journal/EHJ/Thygesen2008}.
Assessment of myocardial viability is essential in the diagnosis and treatment management for patients suffering from MI.
In particular, the position and distribution of myocardial infarct (also known as ``scar") and edema could provide important information for selection of patients and delivery of therapies of MI.
Edema is induced by ischemia and reperfusion, and its size reflects the area of ischemic injury in early acute (per-acute) MI \citep{journal/FCMP/ruder2013}.
Cardiac magnetic resonance (CMR) imaging can be used to determine the effects of acute MI in vivo, as \leireffig{fig:intro:MyoPS} shows. 
For example, the balanced steady-state free precession (bSSFP) sequence can be used to analyze the left ventricular (LV) volume and wall thickness, as it provides a clear LV boundary.
Late gadolinium enhancement (LGE) CMR imaging can visualize infarction, while T2-weighted CMR can depict myocardial edema referring to the area at risk after acute MI. 
To accurately differentiate nonviable infarct myocardium from viable peri-infarct tissues, \citet{journal/Radiology/kidambi2013} defined infarct zone on the 90-day LGE images and peri-infarct zone on the 2-day T2-weighted images acquired from the same patient. 
Therefore, the edema can be divided into two regions of interest around the infarction: the infarct zone and peri-infarct zone, as \leireffig{fig:intro:peri-infarct} (a) shows. 
The task of our challenge is to segment myocardial pathology by combining the three-sequence CMR images from the same patient, assuming the three sequences are aligned prior to pathology segmentation. 
This task is illustrated in \leireffig{fig:intro:MyoPS}.

\subsection{Challenge}\label{intro:challenges}
As manual segmentation is time-consuming and subjective, automatic myocardial pathology segmentation (MyoPS) is highly demanded.
However, automating this segmentation remains challenging, due to the large shape variability of myocardium, indistinguishable boundaries, and the possible poor image quality.
Particularly, there are three challenges for the automatic multi-image-based pathology segmentation.
Firstly, the intensity distribution of the pathological myocardium in LGE and T2 CMR images is heterogeneous.
Secondly, the enhancements of pathologies can be highly variable and complex.
The location, shape and size of infarcts and edemas vary greatly across different patients.
Finally, the misalignment of inter-sequence images introduces new challenges to combine them for the pathology segmentation.

To the best of our knowledge, few work has been reported for MyoPS combining multi-sequence CMR images.
Most of the work only segments single pathology, i.e., scars or edema, based on a single CMR sequence.
This could be due to the difficulty of correcting the misalignment among different sequences.
Therefore, we defined the task of MyoPS where three-sequence CMR images from the same subject were pre-aligned in the challenge event.
This was to mitigate the difficulty of misalignment and data missing \citep{journal/pami/Zhuang2019}, and to encourage the participants to solely focus on the algorithms of MyoPS.

\subsection{Motivation}\label{intro:motivation}

We therefore organized the MyoPS challenge 2020 in conjunction with MICCAI 2020. 
Specifically, the challenge provided three-sequence CMR from 45 subjects and was aimed to encourage the development of new segmentation algorithms which could combine the complementary information from the three CMR sequences.
Twenty-three submissions were evaluated before the deadline, and fifteen teams presented their work at the conference event.
In this paper, we introduce the related information, review the methodologies, and analyze their results in detail.
Our target is to raise interest in studies on pathology segmentation of myocardium combining multi-source images, which has been employed for studies of other organs, such as segmentation of brain tumor \citep{journal/NC/du2016} and prostate cancer \citep{link/I2CVB2016}.

The rest of this paper is structured as follows: 
Section \ref{related work} presents an overview of related work in the previous challenges of MICCAI and their benchmarks, as well as the difficulties and current solutions.
Section \ref{material} provides details of the materials and evaluation framework from the challenge.
Section \ref{methods} summarizes the current methods for MyoPS.
Section \ref{result} describes the results, followed by discussions in Section \ref{discussion}.
Finally, we conclude this work in Section \ref{conclusion}.

\section{Related work}\label{related work}

\subsection{Related challenges and benchmarks} \label{literature:challenges}

\begin{table*}\center
\caption{Summary of previous challenges related to the cardiac segmentation from MICCAI/ ISBI society.
LV: left ventricle; Myo: myocardium; RV: right ventricle;
LA: left atrium; WHBP: whole heart blood pool; WH: whole heart; 
SM: single modality; MM: multi-modality;
MI: myocardial infarction;
MH: myocardial hypertrophy;
ConHD: congenital heart disease;
DCM: dilated cardiomyopathy;
CorHD: coronary heart disease;
AF: atrial fibrillation;
HCM: Hypertrophic cardiomyopathy;
HHD: Hypertensive Heart Disease;
ARV: abnormal right ventricle;
AHS: athlete's heart syndrome;
IHD: ischemic heart disease;
LVNC: left ventricle non-compaction;
$^\ddagger$: multi-center datasets.}
\label{tb:table:challenge}
{\small
\begin{tabular}{ l|lllll} \hline
Challenge	   &  Year & Source & Data info & Target & Pathologies \\
\hline
\citet{journal/ij/radau2009} 		  & 2009 & SM           & 45 bSSFP CMR	            & LV, Myo	  & MI, MH\\
\citet{conf/stacom/Suinesiaputra2011} & 2011 & MM           & 200 bSSFP CMR	            & LV, Myo     & MI \\
\citet{journal/MedIA/Petitjean2015}   & 2012 & SM           & 48 bSSFP CMR	            & RV		  & ConHD \\
\citet{journal/MedIA/karim2016}	   	  & 2012 & SM           & 30 LGE CMR	            & LV scars	  & MI\\
\citet{journal/jcmr/Karim2013}		  & 2013 & SM$^\ddagger$ & 60 LGE CMR                & LA scar	  & AF\\
\citet{journal/tmi/Tobon2015}		  & 2013 & MM           & 30 CT, 30 bSSFP CMR       & LA		  & AF\\
\citet{journal/MedIA/karim2018}	   	  & 2016 & MM           & 10 CT, 10 black-blood CMR & LA wall     & AF\\
\citet{link/HVSMR2016}				  & 2016 & SM           & 20 bSSFP CMR		        & WHBP, Myo	  & ConHD \\
\citet{journal/tmi/bernard2018}	   	  & 2017 & SM           & 150 bSSFP CMR		        & LV, Myo, RV & MI, MH, DCM, abnormal RV\\
\citet{journal/MedIA/zhuang2019}	  & 2017 & MM$^\ddagger$ & 60 CT, 60 bSSFP CMR       & WH          & AF, ConHD, CorHD \\
\citet{journal/MedIA/xiong2020}	      & 2018 & MM           & 150 LGE CMR		        & LA		  & AF \\
\citet{journal/MedIA/zhuang2020}      & 2019 & MM           & 45 bSSFP, LGE, T2 CMR     & LV, Myo, RV & MI \\
\citet{journal/Data/lalande2020}      & 2020 & SM           & 150 LGE                   & LV scars    & MI \\
\citet{journal/TMI/campello2021}      & 2020 & SM$^\ddagger$ & 150 bSSFP CMR             & RV, LV, Myo & \tabincell{l}{HCM, DCM, HHD, ARV,\\  AHS, and IHD} \\
\hline
\end{tabular}}\\
\end{table*}

In recent years, there are many challenges of computational modeling, segmentation and computer-aided diagnosis for cardiovascular problems.
Thanks to those challenges, researchers can develop, test and compare computational algorithms on the same dataset.
\leireftb{tb:table:challenge} presents the recent challenges and public datasets for cardiac segmentation.
One can see that only \citet{journal/MedIA/karim2016} and \citet{journal/jcmr/Karim2013} focused on the LV/ left atrial (LA) scar segmentation from LGE CMR. 
None of them was aimed to combine multi-source images. 
Though there was a challenge with multi-sequence CMR images, it was aimed to segment myocardium from LGE CMR by referring to the training images from other sequences \citep{journal/MedIA/zhuang2020}.
In contrast, MyoPS challenge was aimed to segment and classify the pathology of myocardium combining the complementary information related to the pathology and morphology from the three-sequence CMR, i.e., the bSSFP, T2 and LGE CMR.

\begin{figure*}[thb]\center
	
	\begin{tabular}{@{}cc@{}} 
	\includegraphics[width=0.46\textwidth]{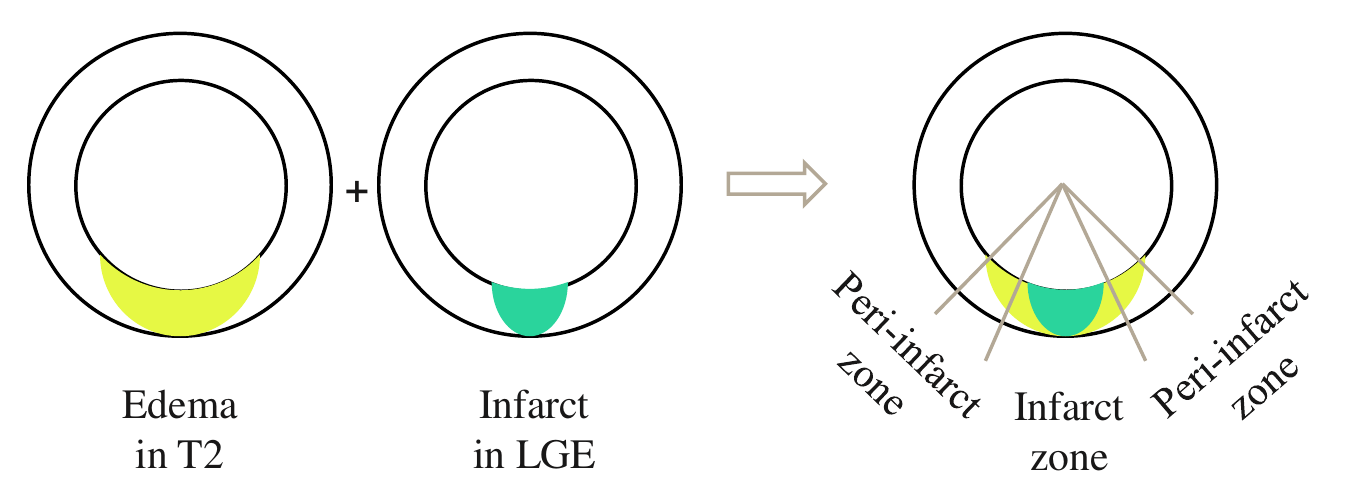}&
	\includegraphics[width=0.49\textwidth, page=1]{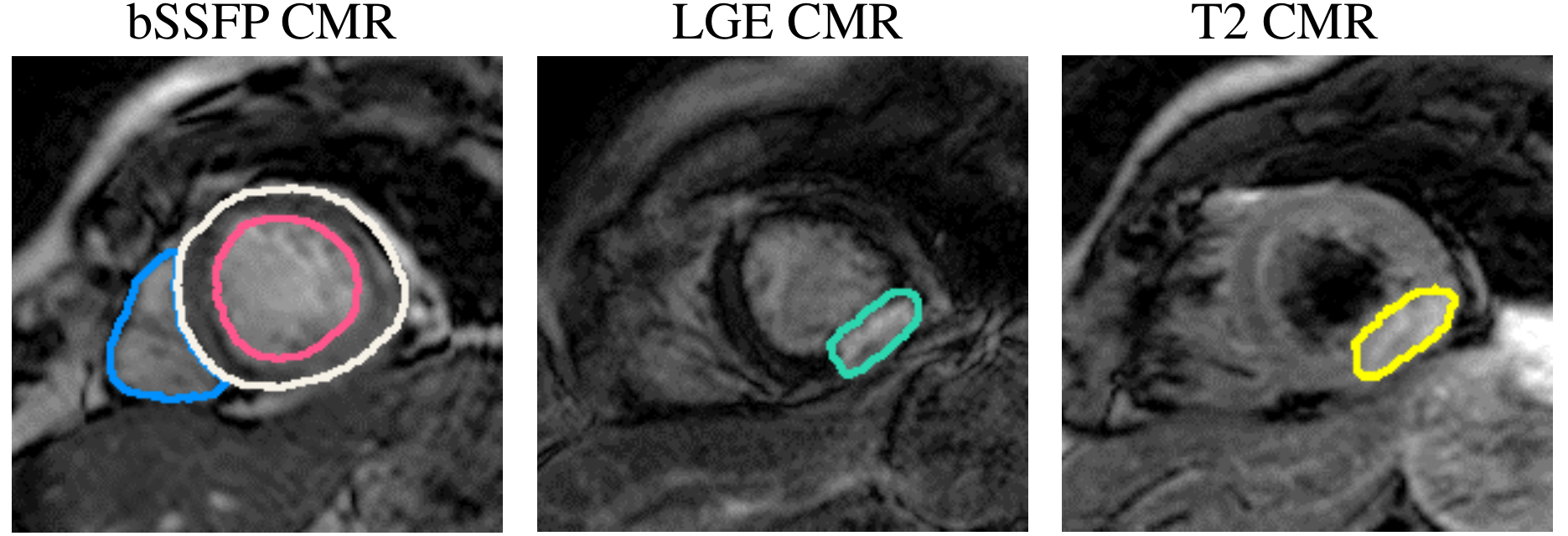}\\
	(a) illustration &(b) from apical slice \\\\[-2ex]
	\includegraphics[width=0.49\textwidth, page=2]{A_figs/fig_MyoPS_image_examples}&
	\includegraphics[width=0.49\textwidth, page=3]{A_figs/fig_MyoPS_image_examples} \\
	(c) from middle slice &(d) from basal slice
	\end{tabular}
	\caption{Illustration of the relationship between myocardial infarct and edema (a), and the exemplary three-sequence CMR images superimposed with contours of gold standard segmentation (b)-(d).}
	\label{fig:intro:peri-infarct}
\end{figure*}

Current pathology segmentation challenges combining multi-modality images mainly came from brain image analysis.
For example, multi-modal brain tumor segmentation (BraTS) challenge was organized in conjunction with the MICCAI 2012-2020 conferences.
The BraTS challenge provided native (T1), post-contrast T1-weighted (T1Gd), T2-weighted (T2), and T2 fluid attenuated inversion recovery (T2-FLAIR) MR images for the brain tumor segmentation.
The first corresponding benchmark study \citep{journal/TMI/menze2014} summarized eleven submitted algorithms, which were all conventional methods, such as fuzzy clustering, level set, and support vector machine.
It found that different algorithms could achieve high performance on a specific sub-region, but no one performed consistently better than the others for all sub-regions.
The next benchmark study \citep{journal/arxiv/bakas2018} of the challenge aimed to assess the state-of-the-art machine learning methods for multi-modal brain tumor segmentation, during BraTS challenge 2012-2018. 
Ischemic stroke lesion segmentation (ISLES) challenge offered at least the set of T1-weighted, T2, diffusion weighted imaging (DWI) and FLAIR MR sequences for each case with a diagnosis of ischemic stroke.
Their benchmark study found that no algorithmic characteristic of any methods was proved better than others, and emphasized the importance of the characteristics of stroke lesion appearances, their evolution and the observed challenges \citep{journal/MedIA/maier2017}.

Other related challenges for other organs include the I2CVB \citep{link/I2CVB2016} and CHAOS  \citep{journal/MedIA/kavur2021}. 
I2CVB provided a multi-parametric MR image dataset, including T2 MR, dynamic contrast enhanced (DCE) MR, DWI MR and MR spectroscopic imaging data, and was aimed for prostate cancer segmentation \citep{link/I2CVB2016}.
CHAOS combined CT and MR images from the abdomen for organ segmentation, including liver, kidneys and spleen \citep{journal/MedIA/kavur2021}. 
To the best of our knowledge, there is still no challenge/ public available dataset on cardiac pathology segmentation combining multi-source images.

\begin{table*} \center
\caption{Summary of current myocardial pathology segmentation algorithms.
    CF: connectivity filtering;
    RG: region growing;
    Error: error in predicted scar/edema percentage;
    RF: random forest;
    RMSE: root mean squared error;
    HD: Hausdorff distance;
    LCE: late contrast enhancement;
    ICC: intraclass correlation coefficient;
    GMM: Gaussian mixture model.
}
\label{tb:table:literature}
{\small
\begin{tabular}{p{3.7cm}|p{2cm}p{1.5cm}p{3.7cm}p{4.8cm}}
\hline
\textbf{Reference} & \textbf{Data} & \textbf{Target(s)} & \textbf{Method} & \textbf{Results} \\
\hline 
\citet{conf/CC/baron2008}          & 22 LGE CMR + T2 CMR & Scar + Edema & Fuzzy clustering & Volume correlation: r$>$0.8 \\
\hline \hline
\citet{journal/MRM/tao2010}        & 20 LGE CMR  & Scar   & Otsu + CF and RG        & \tabincell{l}{Dice = 0.83 ± 0.07 \& 0.79 ± 0.08;\\ Error = 0.0 ± 1.9\% \& 3.8 ± 4.7\%} \\
\hdashline
\citet{journal/QIMS/lu2012}        & 9 LGE CMR    & Scar  & Graph-cuts	            & N/A \\
\hdashline
\citet{journal/JCCT/sandfort2017}  & 34 LCE CT    & Scar  & Adaptive threshold      & \tabincell{l}{Dice = 0.47; \\ ICC (volume/area) = 0.96/0.87} \\
\hdashline
\citet{journal/MedIA/xu2018}       & 165 cine CMR & Scar  & LSTM-RNN + optical flow & \tabincell{l}{Accuracy = 0.95; Kappa = 0.91; \\ Dice = 0.90; RMSE = 0.72 mm;\\  HD = 5.91 mm} \\
\hdashline
\citet{conf/ICPR/kurzendorfer2018} & 30 LGE CMR   & Scar  & Fractal analysis + RF   & Dice = 0.64 ± 0.17 \\
\hdashline
\citet{journal/MRMPBM/moccia2019}  & 30 LGE CMR   & Scar  & FCN                     & \tabincell{l}{Sensitivity = 0.881; Dice = 0.713} \\
\hdashline
\citet{journal/MP/zabihollahy2019} & 34 LGE CMR   & Scar  & CNN                     & \tabincell{l}{Dice = 93.63 ± 2.6\%; \\ Jaccard = 88.13 ± 4.70\%} \\
\hline \hline
\citet{journal/EMBS/kadir2011}     & 16 T2 CMR    & Edema & Morphological filtering + threshold & \tabincell{l}{Volume correlation: r$>$0.8; \\ Error = 9.95± 3.90\%} \\
\hdashline
\citet{journal/JCMR/gao2013}       & 25 T2 CMR    & Edema & Rayleigh-GMM & Dice = 0.74 \\
\hline
\end{tabular}}\\
\end{table*}

\subsection{State-of-the-art myocardial pathology segmentation} \label{literature:methods}

A short overview of previously published algorithms related to MyoPS is presented here, and summarized in \leireftb{tb:table:literature}.
One can see that only \citet{conf/CC/baron2008} segmented both myocardial scar and edema, respectively from LGE CMR and T2 CMR.
Other studies only focus on one of them.
In specific, for scar segmentation the most widespread methods are mainly based on thresholding, such as $n$-SD and full-width-at-half-maximum (FWHM) \citep{journal/MedIA/karim2016,journal/JCCT/sandfort2017}.
It is mostly attribute to the relatively evident intensity contrast between scarring areas and background inside the myocardium.
Instead of simply using thresholding, \citet{journal/MRM/tao2010} combined it with connectivity filtering and region growing.
Other conventional methods were also employed, such as fuzzy clustering \citep{conf/CC/baron2008}, graph-cuts \citep{journal/QIMS/lu2012}, and fractal analysis with random forest \citep{conf/ICPR/kurzendorfer2018}.
Recently, thanks to the great advance in deep learning (DL), \citet{journal/MRMPBM/moccia2019} and \citet{journal/MP/zabihollahy2019} employed fully convolutional networks (FCN) and convolutional neural networks (CNN) for LV scar segmentation.
One can see that most of works extracted scars from LGE CMR or late contrast enhancement CT, where scars are enhanced to distinguish them from non-scarring areas of the myocardium.
However, for the patients with chronic end-stage kidney diseases, the administration of gadolinium contrast agent is dangerous.
Therefore, \citet{journal/MedIA/xu2018} proposed an effective method to directly obtain the position, shape, and size of an infarction area from a raw CMR sequence, i.e., cine CMR.
Compared with LV scar segmentation, there were few works on LV edema segmentation.
The two works listed in \leireftb{tb:table:literature} for edema segmentation were both based on conventional segmentation methods, and no DL-based method was reported, to the best of our knowledge. 
Note that here we only focus on the LV myocardial pathology, and for the literature of another similar topic (i.e., LA myocardial pathology) one could refer to the review \citep{journal/MedIA/li2021}.  

\section{Materials and setup} \label{material}

\subsection{Data}
MyoPS challenge provides 45 paired three-sequence CMR images (bSSFP, LGE and T2 CMR) acquired from the same patient.
The three CMR sequences were all breath-hold, multi-slice, acquired from the cardiac short-axis views using Philips Achieva 1.5T. 
All patients are male with acute MI, and the average age and weight are $56.2 \pm 7.92$ years and $74.4 \pm 5.65$ kg, respectively.
\leireftb{tb:table:dataset} provides the acquisition parameters of the three CMR sequences
\citep{conf/miccai/Zhuang16,journal/pami/Zhuang2019}.
The data acquisition had been anonymized, and approved by the institutional ethics board.
All the data have been pre-processed using the MvMM method \citep{journal/pami/Zhuang2019}, to align the three-sequence images into a common space and to resample them into the same spatial resolution.
\leireffig{fig:intro:peri-infarct} (b)-(d) provide the exemplary images of the three sequences, with contours of gold standard segmentation results superimposed on, from respectively the apical, middle and basal slices.  


\begin{table*}\center
\caption{Image acquisition parameters of the MyoPS challenge data and image parameter after pre-processing. ED: end-diastolic.} 
\label{tb:table:dataset}
{\small
\begin{tabular}{ l|llll} \hline
Sequence  & Imaging type  & TR/TE (ms) & Slice spacing (thickness + gap) & In-plane resolution\\
\hline
LGE	  & T1-weighted              & 3.6/1.8 & 5 mm     & 0.75 $\!\times\!$ 0.75 mm \\
T2    & T2-weighted, black blood & 2000/90 & 12-20 mm &  1.35 $\!\times\!$ 1.35 mm  \\
bSSFP & Cine sequence (ED phase) & 2.7/1.4 & 8-13 mm  &  1.25 $\!\times\!$ 1.25 mm  \\
\hline
\end{tabular}}\\
\end{table*}

For generating the gold standard segmentation, three observers were employed to manually label the LV blood pool, right ventricular (RV) blood pool, and LV myocardium (Myo) from all the three CMR sequences.
In addition, LV myocardial scar and edema were manually delineated from LGE and T2 CMR images, respectively.
The observers were well trained raters who were post-graduate students either in biomedical engineering or medical imaging field.
The manual labeling was performed slice-by-slice using a brush tool in the software ITK-SNAP \citep{journal/Neuroimage/Yushkevich2006}.
All the manual segmentation results were validated by senior experts in cardiac anatomy before used in the construction of gold standard segmentation.
The inter-observer variations of manual scar and edema segmentation in terms of Dice overlap were $0.569\pm0.198$ and $0.701\pm0.168$, respectively. 
The manual segmentation of edema was evidently more consistent between the raters than that of scar segmentation,  probably due to the fact that the regions of edema are generally larger (in terms of size) and less patchy (in terms of shape) from T2 images.

Finally, for the challenge event we split the data into two sets, i.e., the training set, including validation images and consisting of 25 pairs, and the test set composed of 20 pairs.

\subsection{Evaluation metrics} \label{material:evaluation metrics}
Though the labels of LV, RV and Myo were provided, the evaluation of test data only focused on the myocardial pathology segmentation, i.e., scars and edema. 
To evaluate the segmentation accuracy, we calculated the Dice score of scar and edema segmentation separately,
\begin{equation}
	\mathrm{Dice}(V_{\mathrm{seg}},V_{\mathrm{GD}}) = \frac{2\left|V_{\mathrm{seg}} \cap V_{\mathrm{GD}}\right|}{\left|V_{\mathrm{seg}}\right|+\left|V_{\mathrm{GD}}\right|},
\end{equation}
where $V_{\mathrm{GD}}$ and $V_{\mathrm{seg}}$ denote the gold standard and automatic segmentation, respectively.

In addition, we employed three statistical measurements, i.e., Accuracy (ACC), Sensitivity (SEN), and  Specificity (SPE) of the pathology (positive) and healthy myocardium (negative) classification, which are defined as,
\begin{equation}
	\mathrm{ACC}=\frac{TP+TN}{TP+FP+FN+TN}, 
\end{equation}
\begin{equation}
	\mathrm{SEN}=\frac{TP}{TP+FN}, 
\end{equation}
and
\begin{equation}
	\mathrm{SPE}=\frac{TN}{TN+FP},
\end{equation}
where $TP$ and $FP$ respectively stand for the number of pixels of true and false positive myocardial pathologies;
and $TN$ and $FN$ denote the true and false negatives, respectively. 
Note that in this task, the participants were required to solely report the segmentation on pathologies, i.e., to output the pixels labeled as scars or edema in the images, thus the remaining pixels of myocardium not classified as pathologies by an algorithm were then considered as $TN$.
Furthermore, there were cases without pathology, for which we define the sensitivity of a segmentation result to be one, i.e., SEN=1.

\begin{table*}[!t]\center
	\caption{Summary of source code from the participants of MyoPS 2020 challenge.}
	\label{tb:table:team summary}
	{\small
		\begin{tabular}{ l|lll} \hline
			Team    & Code & Reference  \\
			\hline
			UESTC	 & \href{https://github.com/HiLab-git/MyoPS2020}{\textit{https://github.com/HiLab-git/MyoPS2020}} & \citet{conf/MyoPS/zhai2020}       \\
			UBA      & \href{https://github.com/cmartin-isla/MYOPs-challenge-StackedBCDUnet}{\textit{https://github.com/cmartin-isla/MYOPs-challenge-StackedBCDUnet}} & \citet{conf/MyoPS/martin2020}     \\
			NPU      & \href{https://github.com/jianpengz/EfficientSeg}{\textit{https://github.com/jianpengz/EfficientSeg}}  & \citet{conf/MyoPS/zhang2020}      		 \\
			UHW      & \href{https://github.com/chfc-cmi/miccai2020-myops}{\textit{https://github.com/chfc-cmi/miccai2020-myops}} & \citet{conf/MyoPS/ankenbrand2020} 		 \\
			FZU      & \href{https://github.com/kakazxz/myops}{\textit{https://github.com/kakazxz/myops}} & \citet{conf/MyoPS/zhangzhen2020}  		 \\
			NJUST    & \href{https://github.com/JunMa11/MyoPS2020}{\textit{https://github.com/JunMa11/MyoPS2020}} & \citet{conf/MyoPS/ma2020}         		 \\
			CQUPT I  & \href{https://github.com/fly1995/2020MyoPS-MF-DFA-Net}{\textit{https://github.com/fly1995/2020MyoPS-MF-DFA-Net}}  & \citet{conf/MyoPS/Feiyan2020}         	 \\
			LRDE     & \href{https://github.com/Zhaozhou-lrde/myops2020_code}{\textit{https://github.com/Zhaozhou-lrde/myops2020\_code}} & \citet{conf/MyoPS/zhao2020}       		 \\
			CQUPT II & \href{https://github.com/LynnHg/cmsunet}{\textit{https://github.com/LynnHg/cmsunet}} & \citet{conf/MyoPS/li2020} \\
			HNU      & \href{https://github.com/APhun/MyoPS20-HNU}{\textit{https://github.com/APhun/MyoPS20-HNU}} & \citet{conf/MyoPS/liu2020}         \\
			Edin     & \href{https://github.com/falconjhc/MFU-Net}{\textit{https://github.com/falconjhc/MFU-Net}} & \citet{conf/MyoPS/jiang2020}      	 \\
			UBO      & \href{https://github.com/tewodrosweldebirhan/scar_segmentation_myops2020}{\textit{https://github.com/tewodrosweldebirhan/scar\_segmentation\_myops2020}}  & \citet{conf/MyoPS/arega2020}     	 \\
			ITU      & \href{https://github.com/altunokelif/MyoPS2020-CMRsegmentation}{\textit{https://github.com/altunokelif/MyoPS2020-CMRsegmentation}}	& \citet{conf/MyoPS/elif2020}        \\
			UOA      & \href{https://github.com/Voldemort108X/myops20}{\textit{https://github.com/Voldemort108X/myops20}} & \citet{conf/MyoPS/Kumaradevan2020}     	 \\
			\hline
	\end{tabular}}\\
\end{table*}

\subsection{The MyoPS challenge}
\subsubsection{Organization}
We submitted a proposal to the \href{https://www.biomedical-challenges.org}{MICCAI challenge submission system} to apply for our MyoPS challenge.
One can access our challenge proposal in the \href{https://zenodo.org/record/3715932\#.YEXCiWgzZnI}{zenodo website}. 
At the same time, we applied for a \href{https://cmt3.research.microsoft.com/MyoPS2020}{CMT platform} to run this challenge, mainly managing the paper submission.
After preparing all the dataset, we scheduled a timetable for the challenge, including the date of data release, result/ paper submission, associated workshop and result release.
Besides, we designed the task, the distribution of dataset and evaluation metrics.

\subsubsection{Registration and submission}

To access the challenge dataset, researchers were required to sign a data agreement file and return it to the organizers. 
Before the conference, participants can train their model with the training data, and submit their results to the organizers for evaluation.
Each team was allowed to submit their testing results 2 times at most.
After the conference, we have released the encrypted ground truth of test data and corresponding evaluation tool to simplify the evaluation step for subsequent participants.
Therefore, in principle they can evaluate their models unlimited times.

The participants were encouraged to summarize their methods and results by submitting a paper to the 
\href{https://cmt3.research.microsoft.com/MyoPS2020}{CMT-MyoPS platform}.
The format should follow the LNCS style according to the main MICCAI conference guidelines, but we did not constraint the pages.
For the submitted manuscripts, they will be firstly reviewed by the organizers who will ensure the quality of the paper reaches the publication standard. 
Then, each paper will be reviewed by more than two reviewers. 
The review procedure will be double-blinded, similar to the MICCAI submissions. 
Currently, researchers can still download  the MyoPS data and evaluation tool via the \href{www.sdspeople.fudan.edu.cn/zhuangxiahai/0/myops20/}{challenge webpage}. 

\begin{table*}[!t]\center
	\caption{Summary of the benchmarked algorithms. EM: equalization matching; HE: histogram equalization; IN: intensity normalization; RGT: random gamma technique;
		CLAHE: contrast limited adaptive histogram equalization;
		SA: simple augmentation techniques, including random rotation, random flipping, random scaling,  ransom shifting, random cropping, random warping and horizontally flipping.}
	\label{tb:table:method summary}
	{\small 
		\begin{tabular}{p{1.6cm}|p{2.5cm}p{2cm}p{5.8cm}p{4.2cm}} 
			\hline
			Team  & Pre-processing  & Method type & Data augmentation   & Post-processing   \\ 
			\hline
			UESTC	 & crop & two-stage, weighted ensemble & SA & retain the largest connected component     \\\hdashline
			UBA      & crop, IN into [0,1], HE & two-stage  & SA, image synthesis & anatomical constraints, morphological operations \\\hdashline
			NPU      & crop, z-score & end-to-end & SA & remove small isolated regions    		 \\\hdashline
			USTB     & crop, z-score, CLAHE, EM & end-to-end & SA, brightness, contrast shift, non-rigid transformation & remove outliers and disconnected regions \\\hdashline
			UHW      & crop, CLAHE & end-to-end, ensemble & SA, brightness, contrast shift, transformation with simulated MR artifacts &None		 \\\hdashline
			FZU      & crop 	     & two-stage  & SA    & None \\\hdashline
			NJUST    & crop, z-score & two-stage  & SA, brightness, contrast shift	& None	 \\\hdashline
			CQUPT I  & crop, z-score & end-to-end & SA, contrast adjustment, transpose  & None \\\hdashline
			LRDE     & crop, z-score & end-to-end, cascaded & None  & keep the largest connected component  \\\hdashline
			CQUPT II & crop, z-score & end-to-end & SA, mirror, reverse  & None  \\\hdashline
			HNU      & crop, HE, RGT & two-stage  & SA & None \\\hdashline
			Edin     & crop      	 & end-to-end & SA & None \\\hdashline
			UBO      & crop, z-score & end-to-end, cascaded & None & connected component analysis, morphological operations\\\hdashline
			ITU      & crop, IN with zero mean & end-to-end &SA, elastic transformation, image dropping out & None  \\\hdashline
			UOA      & crop, IN      & end-to-end & SA & retain the largest connected component and remove holes \\
			\hline
	\end{tabular}}\\
\end{table*}

\subsubsection{Participants}
As an ongoing event, the challenge has received seventy-six requests of registration before the submission of this manuscript, among which sixty-five teams participated the event before the date of the workshop (Oct 4th, 2020).
Twenty-three submitted results were evaluated before the submission deadline, and fifteen algorithms were included for this benchmark work.
\emph{Note that the team abbreviations in the remaining of this paper refer both to the teams and their corresponding methods, as listed in \leireftb{tb:table:team summary}.}
USTB \citep{conf/MyoPS/yu2020} is not listed here as they did not provide open source code of their algorithm.

\section{Survey of the methods} \label{methods}
For the task of MyoPS, deep learning has attracted the most attention and has also shown great potentials.
Similar to other segmentation tasks, the key to success includes the adoption  of preprocessing, appropriate architecture of networks and loss function, data augmentation, learning strategy, and post-processing.
In this section, we survey the benchmarked methods according to these five aspects.
\leireftb{tb:table:method summary} and \leireftb{tb:table:network info} summarize the key techniques of them, particularly the latter focuses on architecture and training details of the deep neural networks.

\subsection{Preprocessing}
Preprocessing can reduce the complexity of data, and facilitate the models to learn the target knowledge without considering the unnecessary variations.
The widely adopted techniques include cropping regions of interest (ROI) and intensity normalization.

As the pathology to be segmented exists only in the LV, most of the peripheral areas of the background are in fact redundant.
To reduce the complexity from background, all the teams cropped ROIs from the original images prior to the MyoPS.
For example, USTB cropped ROIs of $256\times256$ pixel \citep{conf/MyoPS/yu2020}, and FZU cropped a small ROI and resized into images of $128\times128$ pixel \citep{conf/MyoPS/zhangzhen2020}.
Another method was to perform a coarse segmentation on the images, to localize the position of LV, and then extract ROIs automatically.
For example, UBA adopted a U-Net to predict the myocardial region, and then cropped the smallest bounding box around the myocardium with a small margin of 10 pixels \citep{conf/MyoPS/martin2020}.
As cine CMR presents clear structures of LV while lacking appearance of pathological regions, they chose this modality as the input of the localization U-Net.
Similarly, NJUST used a U-Net to segment the whole LV, and then cropped LV ROI into $112\times112$ from LGE and T2 CMR based on the segmentation results \citep{conf/MyoPS/ma2020}.

Intensity normalization aims to transform the intensity ranges of images into the same one.
Z-score is a common and simple method, which normalizes the data into zero mean and unit standard deviation;
another one is to linearly transform the intensity range of an image into $[0,1]$, which was used by UBA.
More advanced preprocessing involves the application of contrast enhancement to the images.
For example, USTB and UHW employed the method of contrast limited adaptive histogram equalization (CLAHE) \citep{conf/CVGIP/pizer1987}, which is particularly useful for images with low contrast;
and UBA further used histogram equalization on the cropped ROIs to enhance the contrast.
For a summary of all the teams, one can refer to \leireftb{tb:table:method summary} for details.

\begin{sidewaystable*}
\caption{Network architectures and training details of the benchmarked algorithms.
	CE: cross entropy; BCE: binary cross entropy; MI: mutual information; SE: Squeeze-and-Excitation.
	Here, x (*) refers to the number of ensemble models, and Efficient-B1/B2/B3 refer to the EfficientNet with different scales.
	}
\label{tb:table:network info}
{\small
\begin{tabular}{p{1.2cm}|p{3cm}p{2cm}p{2.5cm}p{1.5cm}p{3.5cm}p{1.2cm}p{2cm}p{3.5cm}} 
\hline
Team & Architecture & Ensemble (size) & Batch size & Patch size &  Loss function & Optimizer & Learning rate & Device  \\
\hline
UESTC    & U-Net & x (10) &1 & $160\times 160$ & CE and Dice loss & SGD &6e-3 (decay) & NVIDIA GeForce RTX 2080 Ti \\ \hdashline
UBA      & U-Net, BCDU-Net & x (15) &8 &$256\times 256$  &weighted BCE and Dice loss  & Adam &1e-4 & NVIDIA 1080 GPU\\ \hdashline
NPU      & EfficientNet for encoder, BiFPN for decoder & Efficient-B1/B2/B3 & 64/48/32 for B1/B2/B3 & $288\times 288$& CE, Dice and boundary loss & Adam & 1e-4 (decay) & RTX 2080 Ti \\ \hdashline
USTB     & Dual attention U-Net & None & 8 & $256\times 256$ & Dice loss & SGD &1e-3 (decay)& NVIDIA TITAN RTX \\ \hdashline
UHW      & U-Nets (resnet34 backbone) & x (6) & 12 & $256\times 256$ & CE and Focal loss & Adam &1e-3 (decay) & NVIDIA Tesla K80 \\ \hdashline
FZU      & Channel attention based CNN &None & 16 &$128\times 128$  & Dice loss & Adam & 1e-3 & NVIDIA GeForce RTX 2080 Ti \\	\hdashline
NJUST    & 2D nnU-Net & x (10) & 6 & $112\times 112$ & CE and Dice loss & SGD & 1e-3 & NVIDIA V100\\\hdashline
CQUPT I  & U-Net and a dense connected path & None & 4 & $256\times 256$ & Weighted CE and Dice loss  & Adam & 1e-4 (decay) & NVIDIA Geforce RTX 2080 Ti \\ \hdashline
LRDE     & Cascaded U-Net &None & 1 & $240\times 240$ & BCE & Adam & 1e-4 & NVIDIA Quadro P6000 GPU\\ \hdashline
CQUPT II & Multi-scale U-Net &None & 6 & $256\times 256$ & CE and Dice loss & Adam &  1e-4 (decay) &  NVIDIA Geforce RTX 2080 Ti\\ \hdashline
HNU      & U-Net, attention-based M-shaped network & None &20 & $256\times 256$ & Focal Dice and MSE loss & Adam & 3e-4 & NVIDIA TITAN V GPU\\\hdashline
Edin      & Max-Fusion U-Net & None &4 & $102\times 102$ to $288\times 288$ (96+16i, $1\leq i\leq 12$) & Tversky, focal, and unsupervised reconstruction loss & Adam & 1e-4 & TitanX \\\hdashline
UBO      & Densenet with inception and SE block & None &16 & $350\times 350$ &Logarithmic Dice and region MI loss &Adam &1e-3 & NVIDIA Tesla K80 \\\hdashline
ITU      & Residual U-Net &None & 8 &$256\times 256$  & Dice loss &Adam &1e-3 (decay) & NVIDIA Quadro RTX 6000\\\hdashline
UOA      & A linear encoder and decoder, with a network module consisting of U-Net, Mask-RCNN and U-Net++ & None & 8, 2, 8 for the three components in the network module, respectively &$256\times 256$ & Dice loss for U-Net and U-Net++; classification loss, bounding-box loss and CE loss for Mask-RCNN & Adam &1e-5,1e-3,1e-5 for the three components in the network module, respectively & Tesla P100\\
\hline
\end{tabular}}
\end{sidewaystable*}

\subsection{Architecture and loss function}
The most common architecture in the benchmarked algorithms is U-Net, which extracts multi-scale features and combines them together with a skip connection strategy.
For example, UESTC used U-Net for both coarse and fine segmentation stages.
NPU adopted EfficientNets \citep{conf/ICML/tan2019} as the encoder to extract features from the CMR sequences.
The other useful techniques for feature extraction are the dense connection and attention strategy.
For example, UBA employed the BCDU-Net \citep{conf/ICCV/azad2019} to segment the pathologies.
BCDU-Net is an extension of U-Net and reuses feature maps via dense connections.
USTB embedded a channel attention module and a space attention module at the bottom layer of a U-Net model. 
The former module can selectively emphasize feature association among different channel maps, and the latter captures the long-range dependencies on feature maps.
The effectiveness of these modules was verified in their ablation study.
Moreover, FZU extracted features from the three sequences separately. 
To avoid information redundancy of these features,
 they adopted the channel attention to emphasize the informative features and suppress useless ones.

As to the selection of loss functions, the most commonly used are Dice loss and cross entropy loss.
Nevertheless, boundary loss can also be used to boost the model performance, which is demonstrated in the work of FZU.
This could be attributed to its ability to enforce the model to pay more attention to boundary regions.
Finally, \leireftb{tb:table:network info} provides a summary of model designing and training for the benchmarked methods.

\subsection{Data augmentation}
As the shapes of the myocardium and their pathologies have large variations, the training images could be insufficient, leading to the over-fitting problem of deep learning.
Data augmentation has proven to be effective in improving the generalization ability of  resulting models \citep{conf/TCSVT/takahashi2019}.
We group the augmentation techniques into two categories, i.e., online and offline augmentation. 

The online augmentation includes the random rotation, scaling, shifting, flipping, non-rigid transformations, as well as brightness and contrast adjustment.
For example, USTB adopted the elastic-transform, grid-distortion and optical-distortion to transform the training images non-rigidly.
Experiments showed that this augmentation improved the Dice score by about 8\% for scar segmentation \citep{conf/MyoPS/yu2020}.

The offline augmentation mainly refers to image synthesis.
UBA did a comprehensive synthesis operation \citep{conf/MyoPS/martin2020}.
They utilized the semantic image synthesis with spatially-adaptive normalization (SPADE) method \citep{conf/CVPR/park2019}, to achieve style transfer, pathology rotation, epicardial warping and pathology dilation/ erosion.
Their ablation study demonstrated that these morphological and style transformations could improve the performance significantly.
Interestingly, they found that the style transfer was the most effective, while morphological augmentations, such as the scar and edema dilation and erosion, had limited gains.

\begin{table*} [!t] \center
	\caption{Summary of the quantitative evaluation results of scar and edema segmentation by the fifteen teams.
	Note that column Dice$^\circleddash$  reports the results excluding case $\#207$, which contains no scar; 
	and the average Dice changes from $ 0.614 \pm 0.075 $ to $0.583\pm 0.072$ if case $\#207$ is included. 
	ACC: accuracy; SEN: sensitivity; SPE: specificity.
	}
	\label{tb:result:scar and edema}
	{\normalsize
		\scalebox{0.8}
		{
			\begin{tabular}{  l| l l l l | l l l l *{7}{@{\ \,} l }}\hline
				\multirow{2}*{Team}& \multicolumn{4}{c|}{Scar} & \multicolumn{4}{c}{Edema}\\
				\cline{2-9}
				~      & Dice$^\circleddash$  & {ACC} & {SEN} & {SPE} & Dice  & {ACC} & {SEN} & {SPE} \\
				\hline
UESTC  & \textit{$ 0.708\pm0.191 $} & {$ 0.870\pm0.082  $} & $ 0.737\pm0.185 $ & $ 0.925\pm0.054 $ & {$ 0.731\pm0.109 $} & {$ 0.797\pm0.095 $} & $ 0.724\pm0.134 $ & $ 0.847\pm0.095 $ \\
UBA    & $ 0.701\pm0.189 $ & $ 0.851\pm0.075 $ & {$ 0.791\pm0.175 $} & $ 0.867\pm0.070  $ & $ 0.698\pm0.129 $ & $ 0.762\pm0.102 $ & {$ 0.748\pm0.152 $} & $ 0.770\pm0.099 $\\
NPU    & $ 0.681\pm0.240  $ & $ 0.857\pm0.105 $ & $ 0.734\pm0.253 $ & $ 0.902\pm0.096 $ & $ 0.709\pm0.122 $ & $ 0.777\pm0.112 $ & $ 0.703\pm0.148 $ & $ 0.819\pm0.133 $\\
USTB   & $ 0.668\pm0.255 $ & $ 0.852\pm0.095 $ & $ 0.764\pm0.257 $ & $ 0.872\pm0.093 $ & $ 0.688\pm0.148 $ & $ 0.748\pm0.135 $ & $ 0.741\pm0.164 $ & $ 0.736\pm0.184 $ \\
UHW    & $ 0.652\pm0.195 $ & $ 0.848\pm0.092 $ & $ 0.695\pm0.232 $ & $ 0.891\pm0.108 $ & $ 0.665\pm0.137 $ & $ 0.742\pm0.102 $ & $ 0.722\pm0.193 $ & $ 0.744\pm0.169 $ \\
FZU     & $ 0.627\pm0.215 $ & $ 0.848\pm0.086 $ & $ 0.632\pm0.221 $ & $ 0.931\pm0.043 $ & $ 0.686\pm0.123 $ & $ 0.777\pm0.084 $ & $ 0.663\pm0.151 $ & $ 0.844\pm0.076 $ \\
NJUST  & $ 0.658\pm0.241 $ & $ 0.877\pm0.074 $ & $ 0.642\pm0.269 $ & {$ 0.952\pm0.032 $} & $ 0.599\pm0.200   $ & $ 0.771\pm0.088 $ & $ 0.501\pm0.211 $ & {$ 0.943\pm0.057$}\\
CQUPT I & $ 0.637\pm0.227 $ & $ 0.858\pm0.084 $ & $ 0.626\pm0.223 $ & $ 0.938\pm0.051 $ & $ 0.656\pm0.138 $ & $ 0.766\pm0.096 $ & $ 0.606\pm0.179 $ & $ 0.863\pm0.107 $ \\
LRDE   & $ 0.617\pm0.233 $ & $ 0.809\pm0.142 $ & $ 0.690\pm0.237  $ & $ 0.849\pm0.154 $ & $ 0.639\pm0.141 $ & $ 0.709\pm0.131 $ & $ 0.698\pm0.165 $ & $ 0.716\pm0.199 $\\
CQUPT II & $ 0.612\pm0.237 $ & $ 0.857\pm0.084 $ & $ 0.575\pm0.242 $ & $ 0.951\pm0.048 $ & $ 0.725\pm0.110  $ & $ 0.796\pm0.100   $ & $ 0.709\pm0.156 $ & $ 0.846\pm0.136 $ \\
HNU      & $ 0.581\pm0.243 $ & $ 0.825\pm0.090  $ & $ 0.543\pm0.225 $ & $ 0.923\pm0.060  $ & $ 0.619\pm0.166 $ & $ 0.751\pm0.110  $ & $ 0.544\pm0.190  $ & $ 0.886\pm0.081 $\\
Edin   & $ 0.600\pm0.261   $ & $ 0.836\pm0.106 $ & $ 0.626\pm0.294 $ & $ 0.925\pm0.085 $ & $ 0.603\pm0.182 $ & $ 0.733\pm0.113 $ & $ 0.572\pm0.220  $ & $ 0.843\pm0.127 $ \\
UBO    & $ 0.595\pm0.244 $ & $ 0.806\pm0.096 $ & $ 0.682\pm0.281 $ & $ 0.851\pm0.087 $ & $ 0.664\pm0.150  $ & $ 0.740\pm0.116  $ & $ 0.716\pm0.219 $ & $ 0.760\pm0.142 $  \\
ITU  & $ 0.595\pm0.229 $ & $ 0.824\pm0.098 $ & $ 0.632\pm0.258 $ & $ 0.898\pm0.079 $ & $ 0.612\pm0.160  $ & $ 0.739\pm0.112 $ & $ 0.575\pm0.199 $ & $ 0.838\pm0.102 $\\
UOA    & $ 0.493\pm0.251 $ & $ 0.817\pm0.110  $ & $ 0.453\pm0.273 $ & $ 0.952\pm0.036 $ & $ 0.557\pm0.183 $ & $ 0.718\pm0.127 $ & $ 0.479\pm0.189 $ & $ 0.881\pm0.092 $ \\
\hline \hline
\textit{Average}&$ 0.614 \pm 0.231 $&$ 0.836 \pm 0.096 $& $ 0.643 \pm 0.255 $&  $ 0.904 \pm 0.088 $ &  $ 0.644 \pm 0.153 $ & $ 0.743 \pm 0.112 $& $ 0.645 \pm 0.200 $ &  $ 0.803 \pm 0.148 $ \\
\hline
	\end{tabular} }}\\
\end{table*}

\begin{figure*}[!t]\center
	\includegraphics[width=0.98\textwidth]{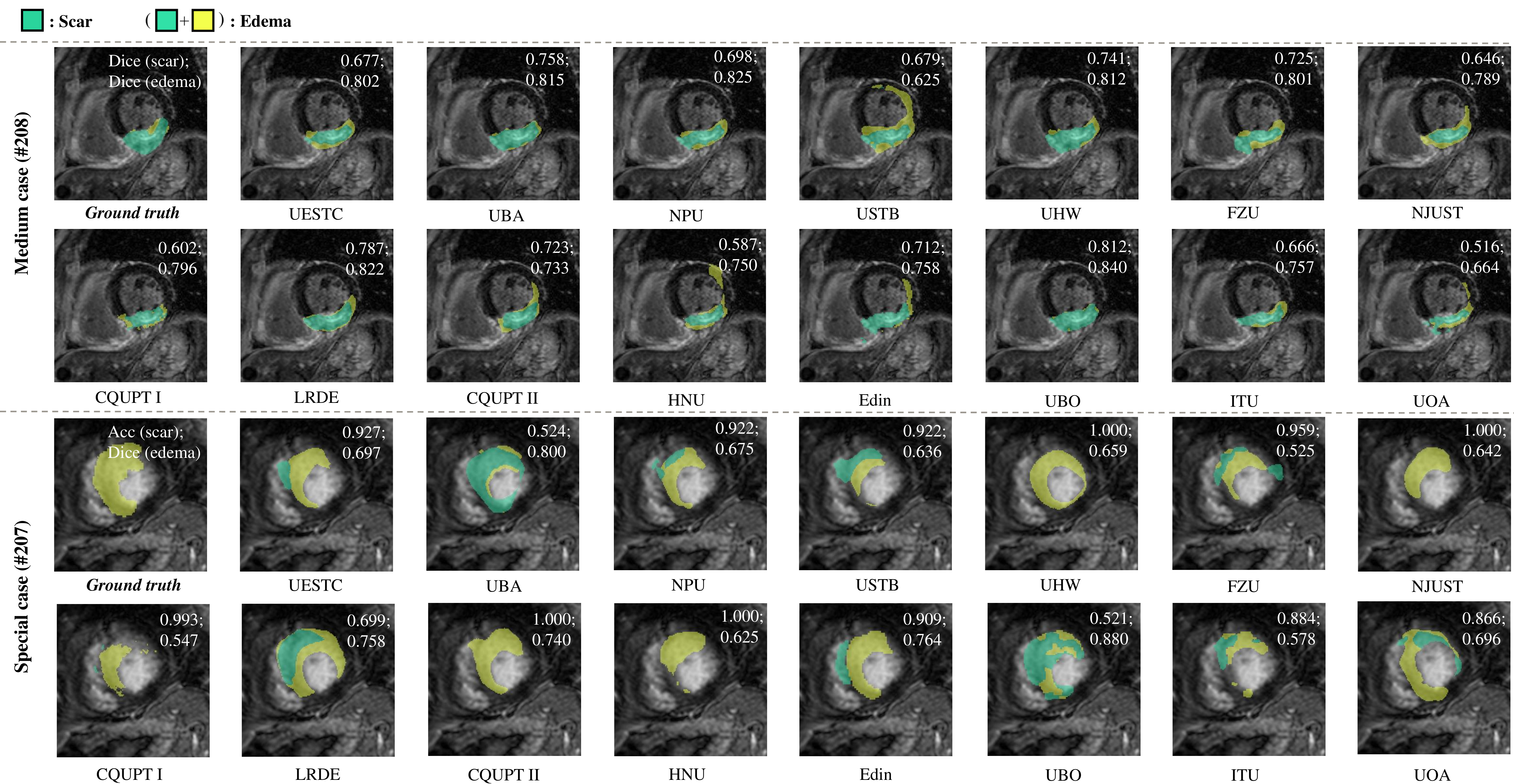}
	\caption{Visualization of the segmentation results of the median and special case by the evaluated methods from each team.
		The median cases were from the test set in terms of mean Dice of scar and edema by the fifteen methods, while the special case contain no scar. Note that the segmentation masks are superimposed on the LGE CMR image which is used for anatomy reference.}
	\label{fig:result:2d_visual}
\end{figure*}

\subsection{Specification of the learning process}
As \leireftb{tb:table:method summary} shows, five teams implemented their works in a two-stage manner (coarse-to-fine), by extracting ROIs on the myocardium prior to a fine process of pathology segmentation.
The others conducted their models in an end-to-end fashion.
In addition, the way of utilizing the extracted ROIs were different, which might explain the discrepancies of their results.
For example, after obtaining the mask including the RV, LV and Myo, UESTC cropped the ROI from all the three sequences and concatenated them using this mask.
They took the concatenation as an input for the final prediction of pathologies.
This strategy can help the segmentation model to take advantage of the extracted knowledge.
Their experiments on the validation dataset have shown advantages of this setting, particularly on the edema with more than 2\% Dice improvement.
Similarly, FZU first learned the mask of LV and Myo, and then did an element-wise multiplication between this mask and the image sequences.
LRDE used three U-Nets to obtain the mask covering the LV and RV, the mask of Myo, and the mask of all three structures from cine and T2 CMR.
These masks were concatenated with the LGE CMR, and then fed into two U-Nets to predict the mask of scar and edema, respectively.
In contrast, UBA solely got the mask of LV, and used it to crop the three sequences.
NJUST obtained a mask of LV and Myo from cine CMR, and used it to crop the other two sequences for prediction of pathologies. 

Another strategy worth mention is model ensemble.
Models trained with different samples or images from different views could learn diverse knowledge.
Their ensemble can make the prediction more robust and accurate.
For example, UESTC employed a 2.5D U-Net to encode in-plane and through-plane information.
Results from the 2.5D U-Net were a weighted average via a 2D U-Net, which outputs the final segmentation.
Experiments showed that this ensemble delivered better results.
UBA adopted a different strategy by generating a number of datasets with synthesized images, and they trained 15 models using different training data.
As their ablation study demonstrated, this ensemble could capture a greater number of non-trivial unconnected components.
Similarly, UHW trained 21 models, but they solely selected the 6 top-performing models for the final aggregation.

\begin{figure*}[!t]\center
    \subfigure {\includegraphics[width=0.92\textwidth]{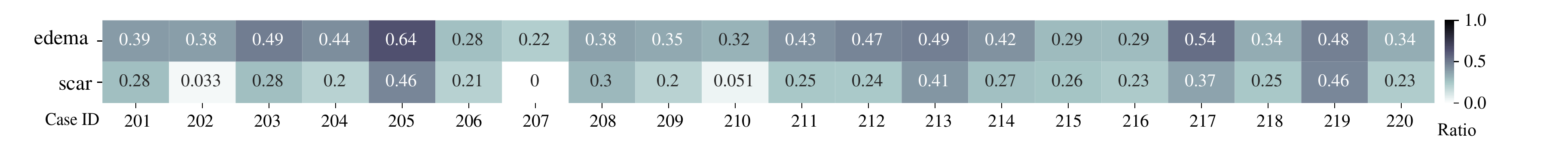}}
    \subfigure {\includegraphics[width=0.92\textwidth]{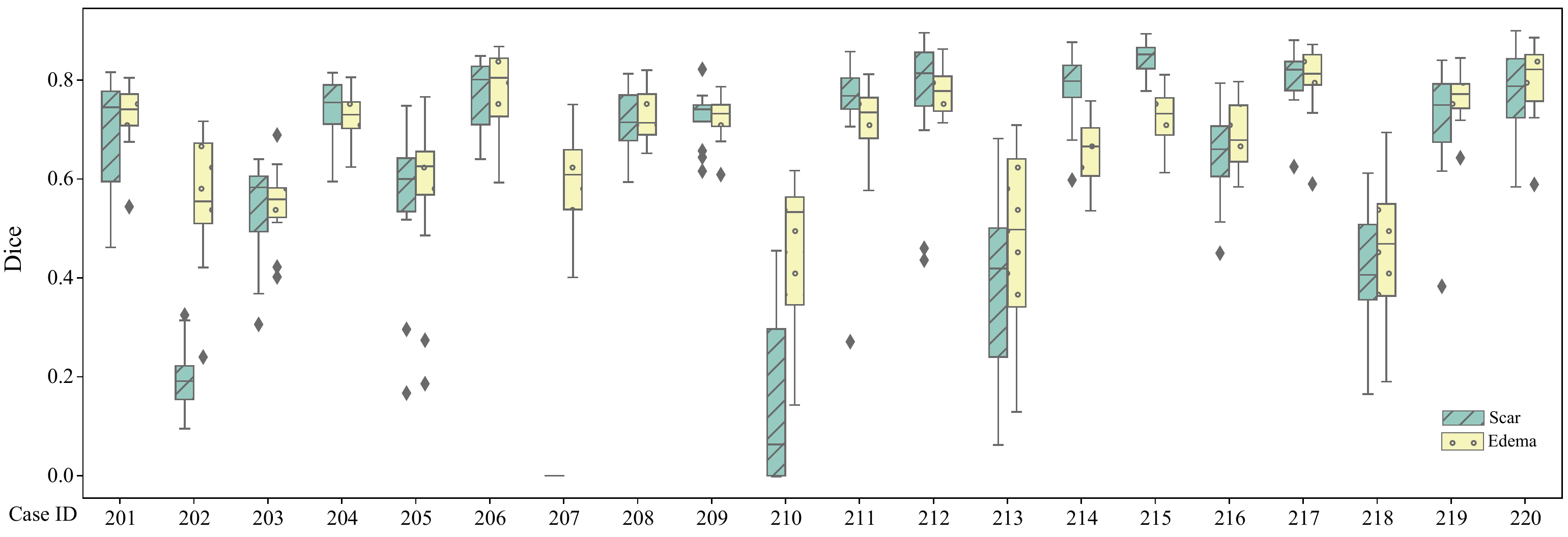}}
    \subfigure {\includegraphics[width=0.92\textwidth]{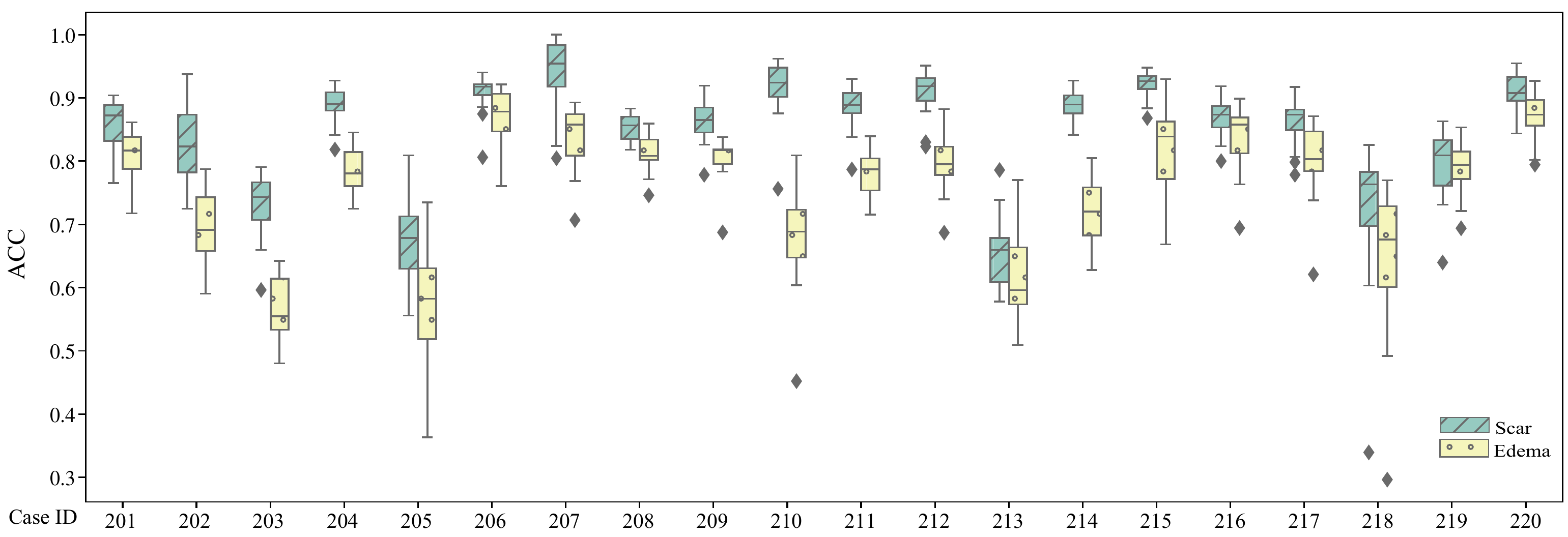}}
   \caption{Ratio of pathologies and the average performance, \textit{i.e.,} Dice and ACC, of the evaluated algorithms on each test case.}
	\label{fig:result:performance_case}
\end{figure*}

\subsection{Post-processing}
Post-processing can be used to remove redundant small patches and refine or regularize the shape of segmentation results to be more realistic.
Among all the benchmarked algorithms, only four conducted post-processing to refine their segmentation results.
Specifically, UBA firstly reconstructed the myocardium mask into a ring shape by extracting a skeleton.
They calculated the distances of pixels around myocardium to the skeleton,
 and those with distance less than a threshold were then categorized into edema.
For the scar segmentation, 3D components smaller than 100 voxels were excluded.
Finally, they did the refinement of the joined edema-scar mask by excluding those 3D components of size smaller than 300 voxels.
The experiments showed that with this post-processing, the Dice of scar segmentation was improved by almost 3\%.
NPU simply removed the small isolated segmentation regions.
USTB discarded the pixels outside the target area as well as unreasonable and unconnected pathological components, and further filled them with adjacent category pixels.
Experiments showed both could improve the pathology segmentation.
UOA employed a post-processing step to solely retain the largest connected component of the predicted LV blood pool and LV epicardium.
Besides, they applied an operation to remove holes that appear inside the foreground masks before the linear decoder.

\section{Results}\label{result}
In this section, we present the results of the evaluated algorithms for comparisons, and then analyze several possible factors that may affect the MyoPS performance.


\subsection{Overall performance}

\leireftb{tb:result:scar and edema} presents the quantitative results of the evaluated algorithms for MyoPS.
The average of mean Dice scores of the evaluated methods are $ 0.614$ and $ 0.644 $ for scar and edema segmentation, respectively; and the average of mean ACC are $ 0.836$ and $ 0.743$, respectively.
In general, the evaluated methods achieved worse performance for scar segmentation than for edema segmentation in terms of Dice, but not in terms of the other three metrics.
In fact, the same metric and value could often refer to different degrees of clinical acceptability for different tasks, depending on the size and shape of the target object and the complexity of form \citep{journal/MedIA/li2020}.
For example, Dice tends to be more sensitive to the small deviations in segmentation for small sparse objects than for large, compact objects, which may explain the better Dice results for edema, which has a larger volume and is less patchy from T2 CMR images.
This conclusion is more evident when we comparing the inter-observer of scar and edema in terms of Dice ($0.569\pm0.198$ vs. $0.701\pm0.168$).
The best Dice scores ($ 0.708\pm0.191 $ and $ 0.731\pm0.109 $ for scar and edema segmentation, respectively) were both achieved by UESTC; 
but the best ACC, SEN and SPE were accomplished respectively by NJUST\&UESTC, UBA\&USTB, and NJUST. 
Interestingly, one algorithm could perform well in one pathology but not necessarily in another, for example CQUPT II excelled in edema segmentation but performed poorly in scar segmentation.


\leireffig{fig:result:2d_visual} visualizes the segmentation results of the middle slice of the special case ($\#207$) and the median case ($\#208$).
Most of the methods achieved good results for the median case ($\#208$), although some contained patchy noises.
Specifically, the results of median case by USTB, HNU, Edin and UOA contain significant amount of outliers of edema, and parts of scars are evidently mis-classified into edema by USTB, FZU and NJUST.
For the special cases ($\#207$), false positives of scar classification were the major errors.
Only UHW, NJUSTM CQUPT II and HUN contained no false positive of scar classification, and thus were evaluated with ACC of 1.000;
UBA and UBO mistook edema as scars, but still obtained high Dice scores for the segmentation of edema which includes both scarring and peri-infarct region.
Nevertheless, this indicates the difficulty of differentiating the scars and peri-infarct regions, which is currently out of the scope of this study.


\leireffig{fig:result:performance_case} provides boxplots of Dice and ACC from  the evaluated algorithms on each test case, where the ratios of pathologies to myocardium are provided for reference. 
One can see that there exist large variations of the performance among different cases in terms of both Dice and ACC.
The position, shape and extent of pathologies all could affect the performance, and will be analyzed in Sections~\ref{result vs. position}, \ref{result vs. shape} and  \ref{result vs. extent}, respectively.
Particularly, the slices without pathology could confuse the algorithms, of which most segment pathology slice-wisely, \textit{i.e.,} an algorithm segments the CMR images slice by slice instead of as a whole volume. 
Note that a slice without pathology could easily induces a Dice score of zero for an algorithm if it mis-classifies even only one pixel, according to its definition.
In the test data, the special case ($\#207$) has no scar  (see \leireffig{fig:result:2d_visual}).

\begin{figure*}[t]\center          
   \includegraphics[width=0.85\textwidth]{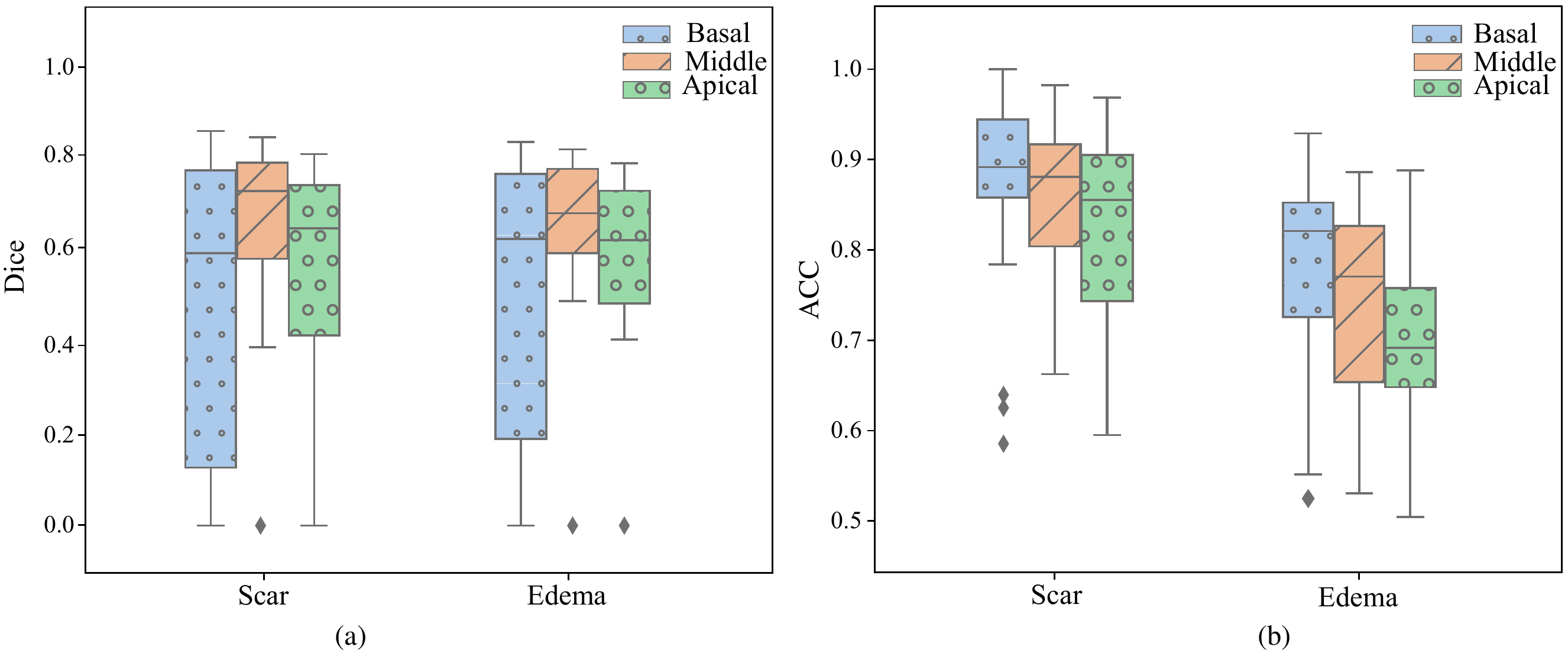} 
   \caption{The boxplots of the average Dice and ACC of pathology segmentation with respect to different slice positions.
   }
	\label{fig:result:pathology_position_map}
\end{figure*}

\begin{figure*}[t]\center
\includegraphics[width=0.98\textwidth]{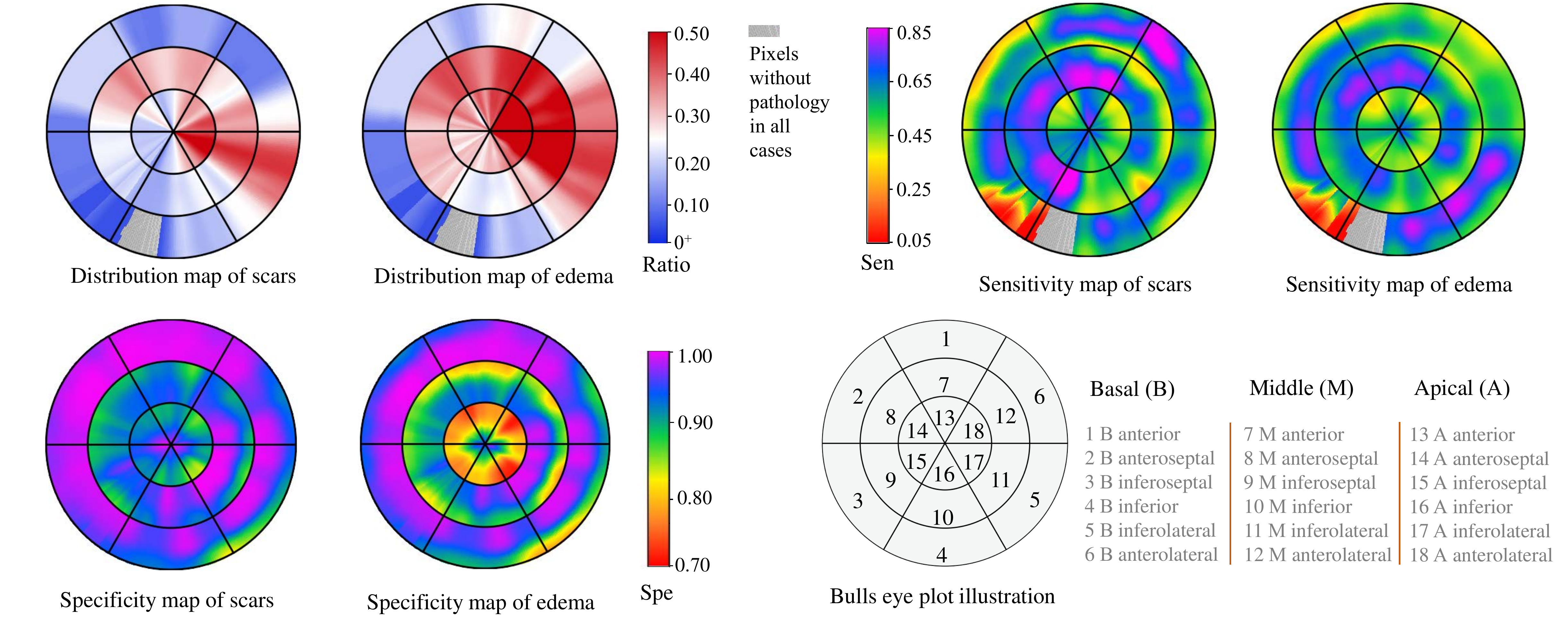}
   \caption{Bulls eye plots of pathology maps and mean segmentation performance with respect to different segments. Note that the grey-colored regions in distribution maps and sensitivity maps indicate none of the cases has pathology in the positions.
   }
	\label{fig:result:pathology_position_bulls_map}
\end{figure*}

\subsection{Performance versus position of pathology} \label{result vs. position}

To analyze the correlation between the performance of MyoPS and the position of pathologies, we first generated the boxplots of Dice and ACC of MyoPS at different slice positions, as shown in \leireffig{fig:result:pathology_position_map}.
It is evident that the results of different slices were different, representing varying types of challenges. 
From the Dice results, which represent overlap of pathologies from two segmentation results, the best performance was observed in the middle slices.
This is reasonable as the ventricles in apical and basal slices usually exhibit more irregular and small-shaped pathologies, which may introduce additional challenges for the segmentation.
Also, the fourth quartile of Basal, \textit{i.e.,} the lowest 25\% of Dice scores were particularly poor. This could be due to less presence of pathologies in basal slices, which is visualized in \leireffig{fig:result:pathology_position_bulls_map}, and a few poor cases inducing particularly low segmentation Dice. 
From the ACC results, whose calculation considers the classification on both of the positives and negatives of pathological segmentation, Basal has higher-valued box plots. This could be again attributed to the rare cases of pathologies occurred in basal slices, which should be discussed below.  
 
\leireffig{fig:result:pathology_position_bulls_map} visualizes the distribution maps of pathologies from the 20 test subjects, and the SEN and SPE maps of MyoPS using 2D bulls eye plots.
As there are various slice numbers among different cases, we normalized the slice positions for each case referring to \citet{journal/European/liu2016} and \citet{conf/FIMH/zhuang2011},
and the maps were averaged from the set of segments from different slices, subjects and different classification results by the benchmarked methods (only for SPE and SEN maps).
From the distribution maps, one can see that  scars mainly occur in the inferolateral regions and anterior segments of middle slices in this test dataset;
and edema extents to more regions of basal slices and almost all segments of apical and middle slices.  

\begin{figure*}[!t]\center
    \includegraphics[width=0.85\textwidth]{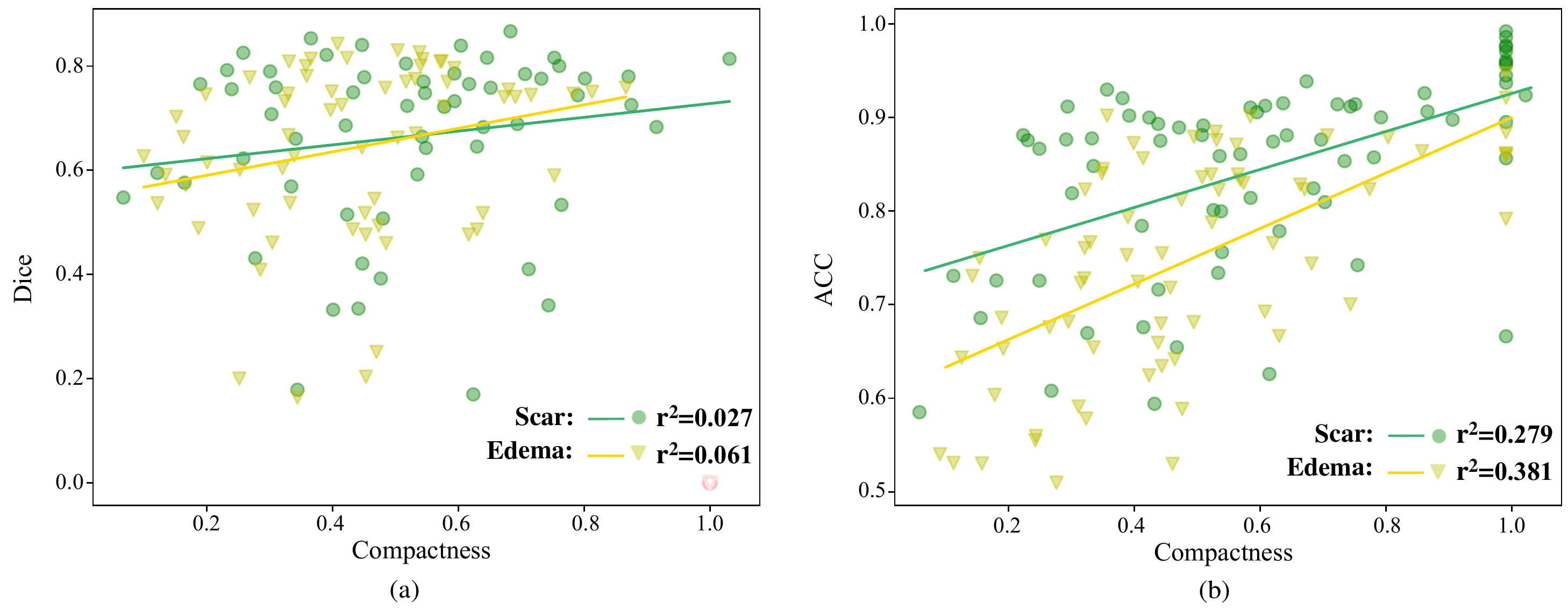}
   \caption{The scatter point plots and correlation between the performance of pathology segmentation with respect to the compactness of the pathologies.  
   Note that the slices without pathologies, indicated by light pink hollow scatter points, are excluded in the computation of correlation with Dice.
   }
\label{fig:result:correlation_compact}
\end{figure*}

From the SEN (true positive rate) and SPE (true negative rate) maps of pathologies, one can observe that the regional values of SPE were generally higher than that of SEN, which could be due to the definition of classification, namely the pixels not segmented as pathologies by an algorithm were regarded as negatives by default.

For scar segmentation, we found that the SEN values were higher in middle septum segments. 
It could be attributed to the good contrast from C0 and T2 for myocardium segmentation of septum, leading to an easier scar segmentation task from LGE myocardium.
By contrast, the low values in SEN maps of both scar and edema are distributed in the area of basal inferoseptal segments, where there should be few cases having pathology, and the models to segment these areas were under trained. 
This explains the particular low Dice in the Dice boxplot of Basal slices in \leireffig{fig:result:pathology_position_map}.
 
Similarly, one can observe from both of the sensitivity maps and specificity maps that the performance of MyoPS on near-endocardium areas was generally better than that on the near-epicardium regions. This could be due to the better contrast in the areas between myocardium and ventricular blood pools than that between myocardium and adjacent tissues (liver and lung) in all the three CMR sequences.   


\subsection{Performance versus shape of pathology} \label{result vs. shape}
\leireffig{fig:result:correlation_compact} presents the correlations between the mean segmentation accuracy (Dice or ACC) and the shape of pathologies.
Here, we employed compactness to quantify the shape of pathologies in a slice, which is defined to the ratio of the area of an object to the area of a circle with the same perimeter \citep{journal/AMC/bogaert2000}.
As a circle is regarded as the object with the most compact shape, the measure normally takes a maximum value of 1 for a circle.
One can see that there are positive correlations between the pathology shape and the performance, which is evident for ACC though marginal for Dice.
This could reveal that the pathologies with asymmetric shapes could be more easily mis-classified by the benchmarked algorithms.


\begin{figure*}[t]\center
    \subfigure {\includegraphics[width=0.85\textwidth]{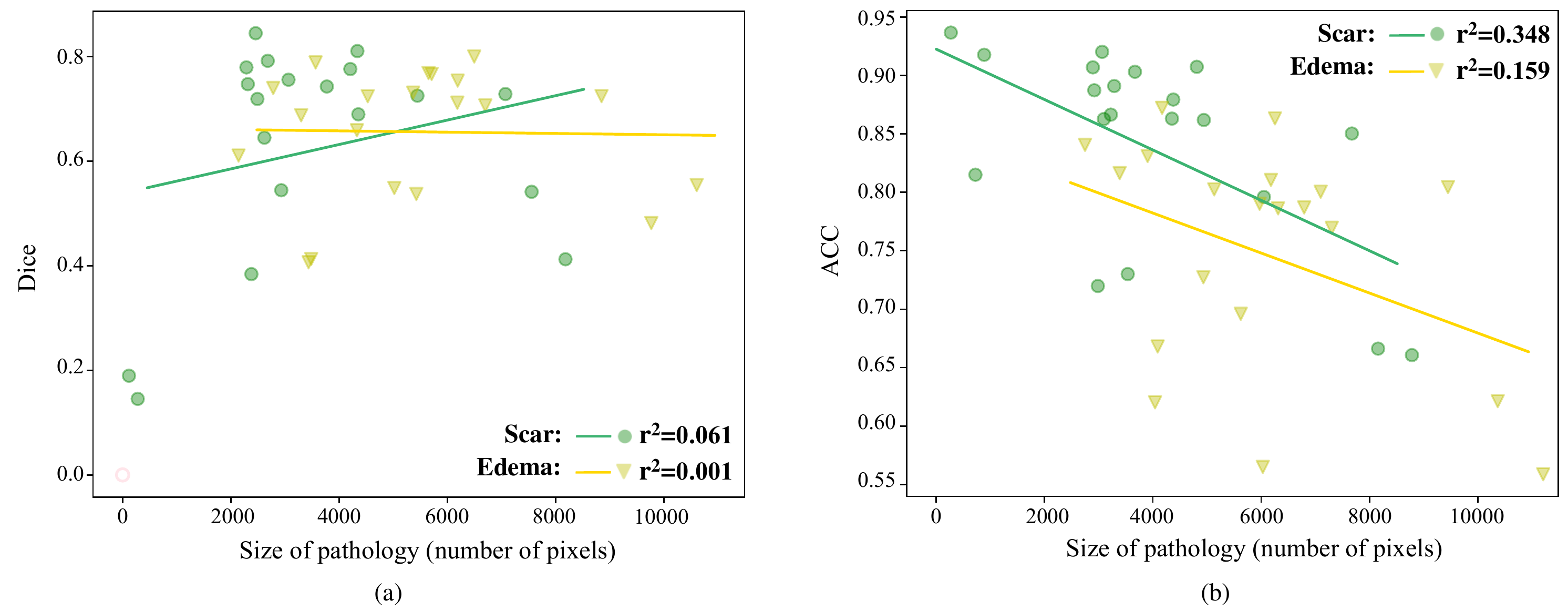}}
    \subfigure {\includegraphics[width=0.85\textwidth]{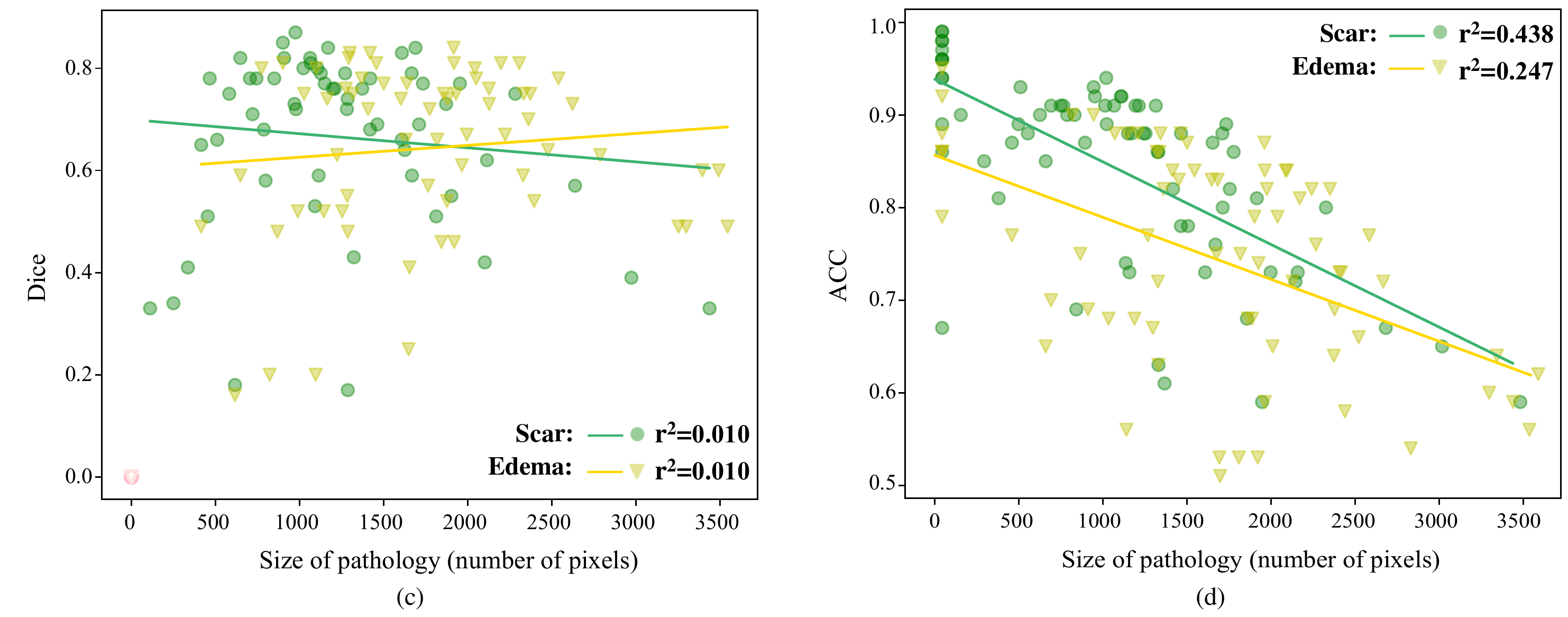}}
   \caption{The scatter point plots and correlation between the performance of pathology segmentation with respect to the size (for Dice score) 
   of the pathologies for patient-wise, in (a) and (b), and slice-wise computation, in (c) and (d).  
   Again, note that the subjects without pathology, indicated by light pink hollow scatter points in (a) and (c), are excluded in the computation of correlation with Dice.}
	\label{fig:result:correlation_extent}
\end{figure*}

\subsection{Performance versus extent of pathology}\label{result vs. extent}

\leireffig{fig:result:correlation_extent} presents the correlations between the mean segmentation accuracy (Dice or ACC) and the extent of pathologies.
We analyzed the correlations from two perspectives, i.e., subject-wise correlation (\leireffig{fig:result:correlation_extent} (a), (b)) and slice-wise correlation (\leireffig{fig:result:correlation_extent} (c), (d)), respectively.
One can see that the ACC values of the pathology segmentation were negatively correlated with extent of pathologies in both the subject-wise and slice-wise studies, but no evident correlation in terms of Dice was observed in either study.
This is reasonable, as pathologies with small size were generally from basal slices in this dataset.
Note that non-pathological myocardium pixels were defined as negatives by default, 
and the cases having large area of negatives tended to have higher ACC values due to this definition. 
In contrast, the variation of pathology sizes did not have an evident influence on the final performance in terms of Dice.

\section{Discussion} \label{discussion}


\subsection{Variation of manual segmentation versus performance and variation of automatic segmentation} 
All the reviewed algorithms were based on supervised learning, so their performance could depend on the quality of labels.
For MyoPS, the inter-observer variability is generally large due to the poor image quality and small volume of targets.
In other words, different experts could offer variable manual segmentation results under the influence of background knowledge and levels of expertise of raters.
To analyze the effect of inter-observer variations on the segmentation performance of automated algorithms, we first performed a correlation analysis between the inter-observer variations and the average performance of all submitted models from participants; 
and we further analyzed the relationship between the inter-observer variations and the inter-participant variations.
Here, inter-observer/ participant variations are defined to the average Dice$^\circleddash$ or Dice scores between different segmentation results.   

\begin{figure*}[t]\center
    \subfigure[] {\includegraphics[width=0.43\textwidth]{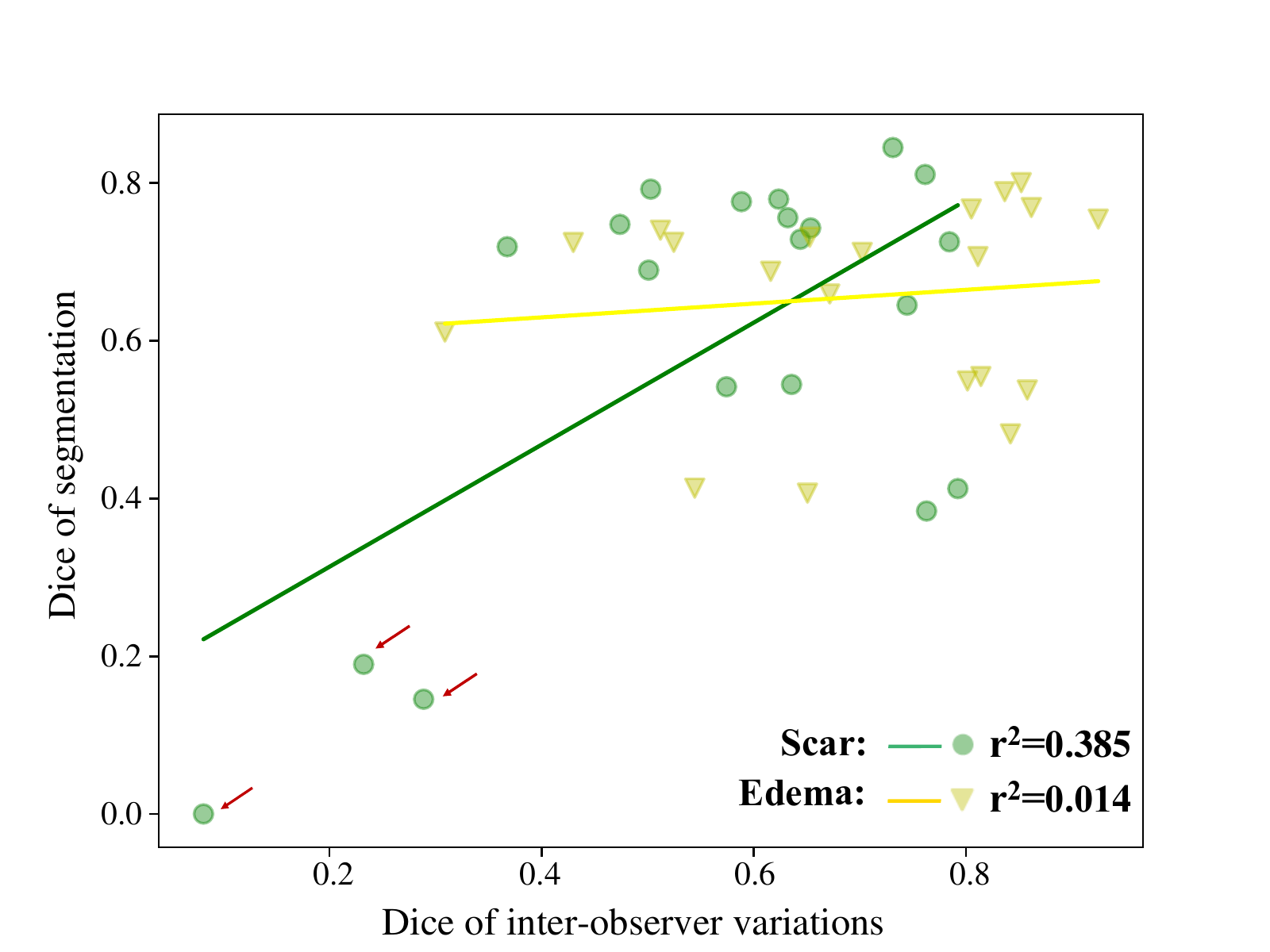}}
    \subfigure[] {\includegraphics[width=0.435\textwidth]{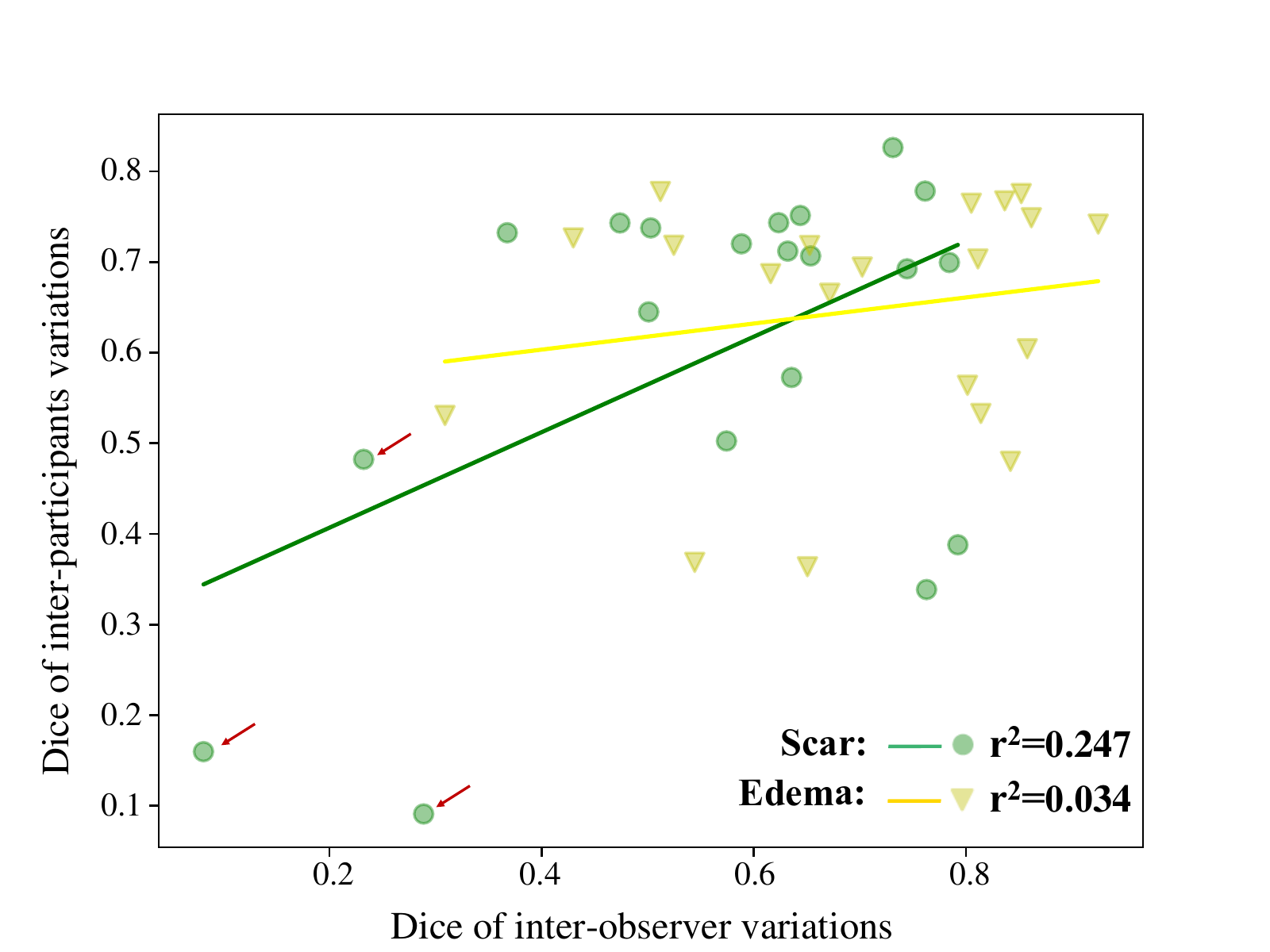}}
   \caption{The scatter point plots and correlations between the inter-observer variations and the average performance in terms of Dice (a), and the inter-participant variations (b). 
   Here, the red arrows identify the cases with large inter-observer variations (low Dice values).
    }
\label{fig:discussion:interob}
\end{figure*}

\leireffig{fig:discussion:interob} presents results of correlation studies. 
The inter-observer variation can be considered as a representation of uncertainty of manual segmentation, which may reveal the difficulties of segmentation.
However, the average Dice scores of the automatic models were not strongly relevant to inter-observer variations. 
Note that the high $r^2$ value for Scar could be attributed to the three special cases highlighted by the red arrows in \leireffig{fig:discussion:interob}.
Similarly, the inter-participant variations can be regarded as the uncertainties of automatic models, which nevertheless had weak correlation to the uncertainties of manual segmentation in this study.

\subsection{Discussion of pre-alignment and MyoPS of CMR}
\begin{figure}[t]\center
    \includegraphics[width=0.43\textwidth]{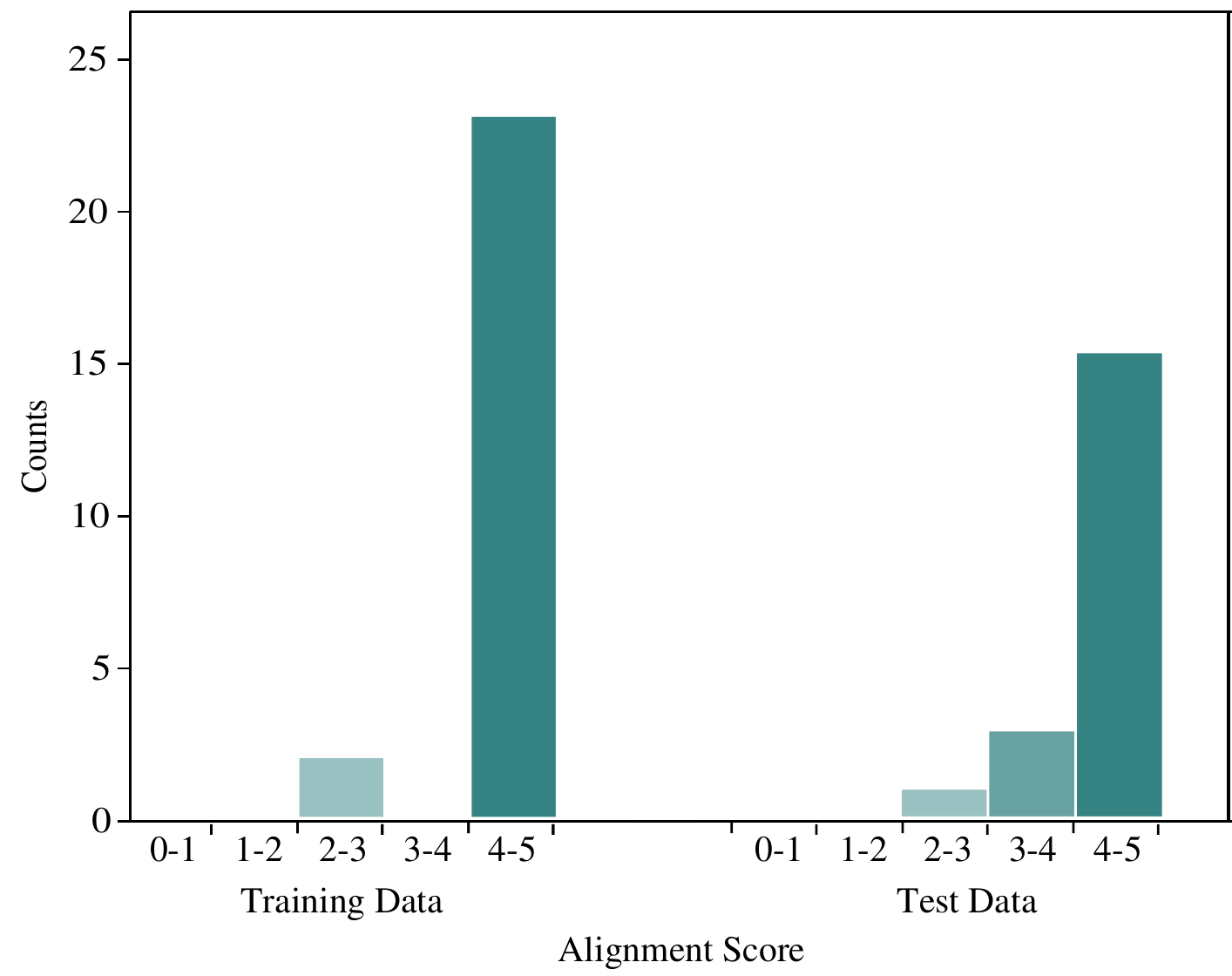}
   \caption{Bar charts of pre-alignment scores in training and test sets of MyoPS dataset.
   }
\label{fig:discussion:alignscore}
\end{figure}

We visually checked each case the alignment result of the three-sequence CMR, and assigned a score, ranging from 0 to 5, to represent the quality of alignment.  
The score of 5 indicates perfect alignment, 1 to 4 denotes misalignment from severe to marginal, and 0 suggests completely failed alignment.
The majority of cases were well aligned, as \leireffig{fig:discussion:alignscore} presents,
and the average scores were $4.50 \pm 0.931$ and $4.68 \pm 0.789$ in the training set and test set, respectively.

One can see that the alignment score is ordinal data in a non-Gaussian distribution.
Therefore, to analyze the effect of pre-alignment on the automatic segmentation, we performed a Spearman's rank correlation analysis \citep{journal/BMJ/sedgwick2014}, on the alignment scores and average Dice$^\circleddash$ (for scar) or Dice (for edema). 
The Spearman coefficient for scar and edema segmentation were respectively -0.207 ($p$-value = 0.740) and -0.143 ($p$-value = 0.252), indicating no evidence of significant relationship between these figures.

The limitation comes from the fact that majority of the cases were well pre-aligned, followed by the segmentation combining the three-sequence MRI. 
Hence, future studies should include the original images without alignments for both training and testing of DL-based models.  
Also, since DL-based method has a great potential to achieve combined computing of simultaneous registration and segmentation. 
Such strategy of combined computing for MyoPS could be further explored in the future.

\subsection{Discussion of evaluation metrics and ranking}
As \leireftb{tb:result:scar and edema} presents, different evaluation metrics could lead to different ranking results for an algorithm or team, indicating potential limitation of the metrics and unfairness of ranking.
Particularly, classification metrics could be misleading for assessment of semantic segmentation.
For example, ACC is sensitive to the volume of targets, and Dice score will fail to act as a metric when the target manual segmentation is none, as discussed in Section \ref{material:evaluation metrics} and \leireffig{fig:result:2d_visual}.
Therefore, we argued that methodology survey and case studies could be more valuable and convincing for benchmark, than ranking the methods according to the figures of evaluation.  

\subsection{Limitation and future prospects} 

There is a gap between technique design and clinical prior knowledge of manual segmentation.
In the clinic, the criteria used to determine the presence of scar/ edema partially relies on the anatomical knowledge of the myocardium.
Specifically, the LV myocardium can be divided into unified 18 segments, displayed on a circumferential polar plot, as shown in \leireffig{fig:result:pathology_position_bulls_map}. 
Despite of variability in the coronary artery blood supply to myocardium, it was believed that every segment can be supplied by specific coronary artery territories \citep{journal/Circulation/american2002}. 
There are three major coronary arteries, each of which supplies its own specific coronary artery territories. 
For example, segments 1, 2, 7, 8, 13 and 14 (left area in bulls-eyes plots) are supplied by the left anterior descending coronary artery. 
With these artery territory knowledge, it is known that the ischemia area (including the scarring/ edema area) generally does not cross two territories, since the successive ischemia is commonly caused by a single vascular occlusion. 
Hence, the predicted area across territory should be penalized. 
However, current methods did not consider this anatomical knowledge of pathology when designing their algorithms.
Therefore, in the future we expect more research on novel methodologies to combine this anatomical knowledge into their framework for more accurate and clinical-related MyoPS results.
Moreover, in this challenge only the short-axis image is used for analysis, while the complementary information from long axis is also crucial in clinical practice for scar localization \citep{journal/JACC/chan2006}. 
In the future, we expect the methods to combine multi-view CMR images for this task.



\section{Conclusion}\label{conclusion}

This paper surveys the submitted works from the MyoPS challenge, which provides 45 sets of three-sequence CMR images. 
Fifteen algorithms were benchmarked for comparisons, and their methodologies and segmentation performance were then analyzed and examined.
To the best of our knowledge, this is the first work to evaluate simultaneous scar and edema segmentation combining multi-source images.
All the benchmarked methods fully utilized the complementary information of the pre-aligned CMR images.
However, none of the benchmarked methods considered the misalignment between the three-sequence CMR images, as they were provided in a pre-aligned format.
We expect more research on simultaneous registration and fusion of multi-source images for pathology segmentation in the future.
Note that the data and evaluation tool continue as ongoing benchmarking resources for researchers.


\section*{Author contributions}
XZ initialized the challenge and provided all the resources;
XZ, LL, FW, SW and XL collected the materials and composed the manuscript.
CM, JZ, YL, ZZ, SZ, MA, HJ, XZ, LW, TA, EA, ZZ, FL, JM, XY, EP, IO, SB, WL, KP, ST, LS, GW, MY, GL, YX, SE were participants of the MyoPS challenge.
The participants described their algorithms and segmentation results for evaluation, and contributed equally to this paper.
All the authors have read and approved the publication of this work.

\section*{Acknowledgement}
This work was funded by the National Natural Science Foundation of China (grant no. 61971142, 62111530195 and 62011540404) and the development fund for Shanghai talents (no. 2020015).

\bibliographystyle{model2-names}
\bibliography{A_ref}

\begin{thebibliography}{64}
\expandafter\ifx\csname natexlab\endcsname\relax\def\natexlab#1{#1}\fi
\providecommand{\url}[1]{\texttt{#1}}
\providecommand{\href}[2]{#2}
\providecommand{\path}[1]{#1}
\providecommand{\DOIprefix}{doi:}
\providecommand{\ArXivprefix}{arXiv:}
\providecommand{\URLprefix}{URL: }
\providecommand{\Pubmedprefix}{pmid:}
\providecommand{\doi}[1]{\href{http://dx.doi.org/#1}{\path{#1}}}
\providecommand{\Pubmed}[1]{\href{pmid:#1}{\path{#1}}}
\providecommand{\bibinfo}[2]{#2}
\ifx\xfnm\relax \def\xfnm[#1]{\unskip,\space#1}\fi
\bibitem[{Ankenbrand et~al.(2020)Ankenbrand, Lohr and
  Schreiber}]{conf/MyoPS/ankenbrand2020}
\bibinfo{author}{Ankenbrand, M.J.}, \bibinfo{author}{Lohr, D.},
  \bibinfo{author}{Schreiber, L.M.}, \bibinfo{year}{2020}.
\newblock \bibinfo{title}{Exploring ensemble applications for multi-sequence
  myocardial pathology segmentation}, in: \bibinfo{booktitle}{Myocardial
  Pathology Segmentation Combining Multi-Sequence CMR Challenge},
  \bibinfo{organization}{Springer}. pp. \bibinfo{pages}{60--67}.
\bibitem[{Arega and Bricq(2020)}]{conf/MyoPS/arega2020}
\bibinfo{author}{Arega, T.W.}, \bibinfo{author}{Bricq, S.},
  \bibinfo{year}{2020}.
\newblock \bibinfo{title}{Automatic myocardial scar segmentation from
  multi-sequence cardiac {MRI} using fully convolutional densenet with
  inception and squeeze-excitation module}, in: \bibinfo{booktitle}{Myocardial
  Pathology Segmentation Combining Multi-Sequence CMR Challenge},
  \bibinfo{organization}{Springer}. pp. \bibinfo{pages}{102--117}.
\bibitem[{Azad et~al.(2019)Azad, Asadi-Aghbolaghi, Fathy and
  Escalera}]{conf/ICCV/azad2019}
\bibinfo{author}{Azad, R.}, \bibinfo{author}{Asadi-Aghbolaghi, M.},
  \bibinfo{author}{Fathy, M.}, \bibinfo{author}{Escalera, S.},
  \bibinfo{year}{2019}.
\newblock \bibinfo{title}{Bi-directional convlstm u-net with densley connected
  convolutions}, in: \bibinfo{booktitle}{Proceedings of the IEEE/CVF
  International Conference on Computer Vision Workshops}, pp.
  \bibinfo{pages}{0--0}.
\bibitem[{Bakas et~al.(2018)Bakas, Reyes, Jakab, Bauer, Rempfler, Crimi,
  Shinohara, Berger, Ha, Rozycki et~al.}]{journal/arxiv/bakas2018}
\bibinfo{author}{Bakas, S.}, \bibinfo{author}{Reyes, M.},
  \bibinfo{author}{Jakab, A.}, \bibinfo{author}{Bauer, S.},
  \bibinfo{author}{Rempfler, M.}, \bibinfo{author}{Crimi, A.},
  \bibinfo{author}{Shinohara, R.T.}, \bibinfo{author}{Berger, C.},
  \bibinfo{author}{Ha, S.M.}, \bibinfo{author}{Rozycki, M.}, et~al.,
  \bibinfo{year}{2018}.
\newblock \bibinfo{title}{Identifying the best machine learning algorithms for
  brain tumor segmentation, progression assessment, and overall survival
  prediction in the {BRATS} challenge}.
\newblock \bibinfo{journal}{arXiv preprint arXiv:1811.02629} .
\bibitem[{Baron et~al.(2008)Baron, Kachenoura, Beygui, Cluze, Grenier, Herment
  and Frouin}]{conf/CC/baron2008}
\bibinfo{author}{Baron, N.}, \bibinfo{author}{Kachenoura, N.},
  \bibinfo{author}{Beygui, F.}, \bibinfo{author}{Cluze, P.},
  \bibinfo{author}{Grenier, P.}, \bibinfo{author}{Herment, A.},
  \bibinfo{author}{Frouin, F.}, \bibinfo{year}{2008}.
\newblock \bibinfo{title}{Quantification of myocardial edema and necrosis
  during acute myocardial infarction}, in: \bibinfo{booktitle}{2008 Computers
  in Cardiology}, \bibinfo{organization}{IEEE}. pp. \bibinfo{pages}{781--784}.
\bibitem[{Bernard et~al.(2018)Bernard, Lalande, Zotti, Cervenansky, Yang, Heng,
  Cetin, Lekadir, Camara, Ballester et~al.}]{journal/tmi/bernard2018}
\bibinfo{author}{Bernard, O.}, \bibinfo{author}{Lalande, A.},
  \bibinfo{author}{Zotti, C.}, \bibinfo{author}{Cervenansky, F.},
  \bibinfo{author}{Yang, X.}, \bibinfo{author}{Heng, P.A.},
  \bibinfo{author}{Cetin, I.}, \bibinfo{author}{Lekadir, K.},
  \bibinfo{author}{Camara, O.}, \bibinfo{author}{Ballester, M.A.G.}, et~al.,
  \bibinfo{year}{2018}.
\newblock \bibinfo{title}{Deep learning techniques for automatic {MRI} cardiac
  multi-structures segmentation and diagnosis: is the problem solved?}
\newblock \bibinfo{journal}{IEEE Transactions on Medical Imaging}
  \bibinfo{volume}{37}, \bibinfo{pages}{2514--2525}.
\bibitem[{Bogaert et~al.(2000)Bogaert, Rousseau, Van~Hecke and
  Impens}]{journal/AMC/bogaert2000}
\bibinfo{author}{Bogaert, J.}, \bibinfo{author}{Rousseau, R.},
  \bibinfo{author}{Van~Hecke, P.}, \bibinfo{author}{Impens, I.},
  \bibinfo{year}{2000}.
\newblock \bibinfo{title}{Alternative area-perimeter ratios for measurement of
  2d shape compactness of habitats}.
\newblock \bibinfo{journal}{Applied Mathematics and Computation}
  \bibinfo{volume}{111}, \bibinfo{pages}{71--85}.
\bibitem[{Campello et~al.(2021)Campello, Gkontra, Izquierdo, Mart{\'\i}n-Isla,
  Sojoudi, Full, Maier-Hein, Zhang, He, Ma et~al.}]{journal/TMI/campello2021}
\bibinfo{author}{Campello, V.M.}, \bibinfo{author}{Gkontra, P.},
  \bibinfo{author}{Izquierdo, C.}, \bibinfo{author}{Mart{\'\i}n-Isla, C.},
  \bibinfo{author}{Sojoudi, A.}, \bibinfo{author}{Full, P.M.},
  \bibinfo{author}{Maier-Hein, K.}, \bibinfo{author}{Zhang, Y.},
  \bibinfo{author}{He, Z.}, \bibinfo{author}{Ma, J.}, et~al.,
  \bibinfo{year}{2021}.
\newblock \bibinfo{title}{Multi-centre, multi-vendor and multi-disease cardiac
  segmentation: The {M\&M}s challenge}.
\newblock \bibinfo{journal}{IEEE Transactions on Medical Imaging} .
\bibitem[{Cerqueira et~al.(2002)Cerqueira, Weissman, Dilsizian, Jacobs, Kaul,
  Laskey, Pennell, Rumberger, Ryan et~al.}]{journal/Circulation/american2002}
\bibinfo{author}{Cerqueira, M.D.}, \bibinfo{author}{Weissman, N.J.},
  \bibinfo{author}{Dilsizian, V.}, \bibinfo{author}{Jacobs, A.K.},
  \bibinfo{author}{Kaul, S.}, \bibinfo{author}{Laskey, W.K.},
  \bibinfo{author}{Pennell, D.J.}, \bibinfo{author}{Rumberger, J.A.},
  \bibinfo{author}{Ryan, T.}, et~al., \bibinfo{year}{2002}.
\newblock \bibinfo{title}{Standardized myocardial segmentation and nomenclature
  for tomographic imaging of the heart: a statement for healthcare
  professionals from the cardiac imaging committee of the council on clinical
  cardiology of the american heart association}.
\newblock \bibinfo{journal}{Circulation} \bibinfo{volume}{105},
  \bibinfo{pages}{539--542}.
\bibitem[{Chan et~al.(2006)Chan, Hanekom, Wong, Leano, Cho and
  Marwick}]{journal/JACC/chan2006}
\bibinfo{author}{Chan, J.}, \bibinfo{author}{Hanekom, L.},
  \bibinfo{author}{Wong, C.}, \bibinfo{author}{Leano, R.},
  \bibinfo{author}{Cho, G.Y.}, \bibinfo{author}{Marwick, T.H.},
  \bibinfo{year}{2006}.
\newblock \bibinfo{title}{Differentiation of subendocardial and transmural
  infarction using two-dimensional strain rate imaging to assess short-axis and
  long-axis myocardial function}.
\newblock \bibinfo{journal}{Journal of the American College of Cardiology}
  \bibinfo{volume}{48}, \bibinfo{pages}{2026--2033}.
\bibitem[{Du et~al.(2016)Du, Li, Lu and Xiao}]{journal/NC/du2016}
\bibinfo{author}{Du, J.}, \bibinfo{author}{Li, W.}, \bibinfo{author}{Lu, K.},
  \bibinfo{author}{Xiao, B.}, \bibinfo{year}{2016}.
\newblock \bibinfo{title}{An overview of multi-modal medical image fusion}.
\newblock \bibinfo{journal}{Neurocomputing} \bibinfo{volume}{215},
  \bibinfo{pages}{3--20}.
\bibitem[{Elif and Ilkay(2020)}]{conf/MyoPS/elif2020}
\bibinfo{author}{Elif, A.}, \bibinfo{author}{Ilkay, O.}, \bibinfo{year}{2020}.
\newblock \bibinfo{title}{Accurate myocardial pathology segmentation with
  residual u-net}, in: \bibinfo{booktitle}{Myocardial Pathology Segmentation
  Combining Multi-Sequence CMR Challenge}, \bibinfo{organization}{Springer}.
  pp. \bibinfo{pages}{128--137}.
\bibitem[{Gao et~al.(2013)Gao, Kadir, Payne, Soraghan and
  Berry}]{journal/JCMR/gao2013}
\bibinfo{author}{Gao, H.}, \bibinfo{author}{Kadir, K.}, \bibinfo{author}{Payne,
  A.R.}, \bibinfo{author}{Soraghan, J.}, \bibinfo{author}{Berry, C.},
  \bibinfo{year}{2013}.
\newblock \bibinfo{title}{Highly automatic quantification of myocardial oedema
  in patients with acute myocardial infarction using bright blood {T}2-weighted
  {CMR}}.
\newblock \bibinfo{journal}{Journal of Cardiovascular Magnetic Resonance}
  \bibinfo{volume}{15}, \bibinfo{pages}{1--12}.
\bibitem[{Jiang et~al.(2020)Jiang, Wang, Chartsias and
  Tsaftaris}]{conf/MyoPS/jiang2020}
\bibinfo{author}{Jiang, H.}, \bibinfo{author}{Wang, C.},
  \bibinfo{author}{Chartsias, A.}, \bibinfo{author}{Tsaftaris, S.A.},
  \bibinfo{year}{2020}.
\newblock \bibinfo{title}{Max-fusion u-net for multi-modal pathology
  segmentation with attention and dynamic resampling}, in:
  \bibinfo{booktitle}{Myocardial Pathology Segmentation Combining
  Multi-Sequence CMR Challenge}, \bibinfo{organization}{Springer}. pp.
  \bibinfo{pages}{68--81}.
\bibitem[{Kadir et~al.(2011)Kadir, Gao, Payne, Soraghan and
  Berry}]{journal/EMBS/kadir2011}
\bibinfo{author}{Kadir, K.}, \bibinfo{author}{Gao, H.}, \bibinfo{author}{Payne,
  A.}, \bibinfo{author}{Soraghan, J.}, \bibinfo{author}{Berry, C.},
  \bibinfo{year}{2011}.
\newblock \bibinfo{title}{Automatic quantification and 3d visualisation of
  edema in cardiac {MRI}}, in: \bibinfo{booktitle}{2011 Annual International
  Conference of the IEEE Engineering in Medicine and Biology Society},
  \bibinfo{organization}{IEEE}. pp. \bibinfo{pages}{8021--8024}.
\bibitem[{Karim et~al.(2016)Karim, Bhagirath, Claus, Housden, Chen,
  Karimaghaloo, Sohn, Rodr{\'\i}guez, Vera, Alb{\`a}
  et~al.}]{journal/MedIA/karim2016}
\bibinfo{author}{Karim, R.}, \bibinfo{author}{Bhagirath, P.},
  \bibinfo{author}{Claus, P.}, \bibinfo{author}{Housden, R.J.},
  \bibinfo{author}{Chen, Z.}, \bibinfo{author}{Karimaghaloo, Z.},
  \bibinfo{author}{Sohn, H.M.}, \bibinfo{author}{Rodr{\'\i}guez, L.L.},
  \bibinfo{author}{Vera, S.}, \bibinfo{author}{Alb{\`a}, X.}, et~al.,
  \bibinfo{year}{2016}.
\newblock \bibinfo{title}{Evaluation of state-of-the-art segmentation
  algorithms for left ventricle infarct from late gadolinium enhancement {MR}
  images}.
\newblock \bibinfo{journal}{Medical Image Analysis} \bibinfo{volume}{30},
  \bibinfo{pages}{95--107}.
\bibitem[{Karim et~al.(2018)Karim, Blake, Inoue, Tao, Jia, Housden, Bhagirath,
  Duval, Varela, Behar et~al.}]{journal/MedIA/karim2018}
\bibinfo{author}{Karim, R.}, \bibinfo{author}{Blake, L.E.},
  \bibinfo{author}{Inoue, J.}, \bibinfo{author}{Tao, Q.}, \bibinfo{author}{Jia,
  S.}, \bibinfo{author}{Housden, R.J.}, \bibinfo{author}{Bhagirath, P.},
  \bibinfo{author}{Duval, J.L.}, \bibinfo{author}{Varela, M.},
  \bibinfo{author}{Behar, J.}, et~al., \bibinfo{year}{2018}.
\newblock \bibinfo{title}{Algorithms for left atrial wall segmentation and
  thickness--evaluation on an open-source {CT} and {MRI} image database}.
\newblock \bibinfo{journal}{Medical Image Analysis} \bibinfo{volume}{50},
  \bibinfo{pages}{36--53}.
\bibitem[{Karim et~al.(2013)Karim, Housden, Balasubramaniam, Chen, Perry,
  Uddin, Al-Beyatti, Palkhi, Acheampong, Obom et~al.}]{journal/jcmr/Karim2013}
\bibinfo{author}{Karim, R.}, \bibinfo{author}{Housden, R.J.},
  \bibinfo{author}{Balasubramaniam, M.}, \bibinfo{author}{Chen, Z.},
  \bibinfo{author}{Perry, D.}, \bibinfo{author}{Uddin, A.},
  \bibinfo{author}{Al-Beyatti, Y.}, \bibinfo{author}{Palkhi, E.},
  \bibinfo{author}{Acheampong, P.}, \bibinfo{author}{Obom, S.}, et~al.,
  \bibinfo{year}{2013}.
\newblock \bibinfo{title}{Evaluation of current algorithms for segmentation of
  scar tissue from late gadolinium enhancement cardiovascular magnetic
  resonance of the left atrium: an open-access grand challenge}.
\newblock \bibinfo{journal}{Journal of Cardiovascular Magnetic Resonance}
  \bibinfo{volume}{15}, \bibinfo{pages}{105}.
\bibitem[{Kavur et~al.(2021)Kavur, Gezer, Bar{\i}{\c{s}}, Aslan, Conze, Groza,
  Pham, Chatterjee, Ernst, {\"O}zkan et~al.}]{journal/MedIA/kavur2021}
\bibinfo{author}{Kavur, A.E.}, \bibinfo{author}{Gezer, N.S.},
  \bibinfo{author}{Bar{\i}{\c{s}}, M.}, \bibinfo{author}{Aslan, S.},
  \bibinfo{author}{Conze, P.H.}, \bibinfo{author}{Groza, V.},
  \bibinfo{author}{Pham, D.D.}, \bibinfo{author}{Chatterjee, S.},
  \bibinfo{author}{Ernst, P.}, \bibinfo{author}{{\"O}zkan, S.}, et~al.,
  \bibinfo{year}{2021}.
\newblock \bibinfo{title}{Chaos challenge-combined ({CT-MR}) healthy abdominal
  organ segmentation}.
\newblock \bibinfo{journal}{Medical Image Analysis} \bibinfo{volume}{69},
  \bibinfo{pages}{101950}.
\bibitem[{Kidambi et~al.(2013)Kidambi, Mather, Swoboda, Motwani, Fairbairn,
  Greenwood and Plein}]{journal/Radiology/kidambi2013}
\bibinfo{author}{Kidambi, A.}, \bibinfo{author}{Mather, A.N.},
  \bibinfo{author}{Swoboda, P.}, \bibinfo{author}{Motwani, M.},
  \bibinfo{author}{Fairbairn, T.A.}, \bibinfo{author}{Greenwood, J.P.},
  \bibinfo{author}{Plein, S.}, \bibinfo{year}{2013}.
\newblock \bibinfo{title}{Relationship between myocardial edema and regional
  myocardial function after reperfused acute myocardial infarction: an {MR}
  imaging study}.
\newblock \bibinfo{journal}{Radiology} \bibinfo{volume}{267},
  \bibinfo{pages}{701--708}.
\bibitem[{Kurzendorfer et~al.(2018)Kurzendorfer, Breininger, Steidl, Brost,
  Forman and Maier}]{conf/ICPR/kurzendorfer2018}
\bibinfo{author}{Kurzendorfer, T.}, \bibinfo{author}{Breininger, K.},
  \bibinfo{author}{Steidl, S.}, \bibinfo{author}{Brost, A.},
  \bibinfo{author}{Forman, C.}, \bibinfo{author}{Maier, A.},
  \bibinfo{year}{2018}.
\newblock \bibinfo{title}{Myocardial scar segmentation in {LGE-MRI} using
  fractal analysis and random forest classification}, in:
  \bibinfo{booktitle}{2018 24th International Conference on Pattern Recognition
  (ICPR)}, \bibinfo{organization}{IEEE}. pp. \bibinfo{pages}{3168--3173}.
\bibitem[{Lalande et~al.(2020)Lalande, Chen, Decourselle, Qayyum, Pommier,
  Lorgis, de~la Rosa, Cochet, Cottin, Ginhac et~al.}]{journal/Data/lalande2020}
\bibinfo{author}{Lalande, A.}, \bibinfo{author}{Chen, Z.},
  \bibinfo{author}{Decourselle, T.}, \bibinfo{author}{Qayyum, A.},
  \bibinfo{author}{Pommier, T.}, \bibinfo{author}{Lorgis, L.},
  \bibinfo{author}{de~la Rosa, E.}, \bibinfo{author}{Cochet, A.},
  \bibinfo{author}{Cottin, Y.}, \bibinfo{author}{Ginhac, D.}, et~al.,
  \bibinfo{year}{2020}.
\newblock \bibinfo{title}{Emidec: a database usable for the automatic
  evaluation of myocardial infarction from delayed-enhancement cardiac {MRI}}.
\newblock \bibinfo{journal}{Data} \bibinfo{volume}{5}, \bibinfo{pages}{89}.
\bibitem[{Li and Li(2020)}]{conf/MyoPS/Feiyan2020}
\bibinfo{author}{Li, F.}, \bibinfo{author}{Li, W.}, \bibinfo{year}{2020}.
\newblock \bibinfo{title}{Dual-path feature aggregation network combined
  multi-layer fusion for myocardial pathology segmentation with multi-sequence
  cardiac {MR}}, in: \bibinfo{booktitle}{Myocardial Pathology Segmentation
  Combining Multi-Sequence CMR Challenge}, \bibinfo{organization}{Springer}.
  pp. \bibinfo{pages}{146--158}.
\bibitem[{Li et~al.(2020a)Li, Udupa, Tong, Wang and
  Torigian}]{journal/MedIA/li2020}
\bibinfo{author}{Li, J.}, \bibinfo{author}{Udupa, J.K.}, \bibinfo{author}{Tong,
  Y.}, \bibinfo{author}{Wang, L.}, \bibinfo{author}{Torigian, D.A.},
  \bibinfo{year}{2020}a.
\newblock \bibinfo{title}{Linsem: Linearizing segmentation evaluation metrics
  for medical images}.
\newblock \bibinfo{journal}{Medical Image Analysis} \bibinfo{volume}{60},
  \bibinfo{pages}{101601}.
\bibitem[{Li et~al.(2021)Li, Zimmer, Schnabel and
  Zhuang}]{journal/MedIA/li2021}
\bibinfo{author}{Li, L.}, \bibinfo{author}{Zimmer, V.A.},
  \bibinfo{author}{Schnabel, J.A.}, \bibinfo{author}{Zhuang, X.},
  \bibinfo{year}{2021}.
\newblock \bibinfo{title}{Medical image analysis on left atrial {LGE MRI} for
  atrial fibrillation studies: A review}.
\newblock \bibinfo{journal}{arXiv preprint arXiv:2106.09862} .
\bibitem[{Li et~al.(2020b)Li, Wang and Qin}]{conf/MyoPS/li2020}
\bibinfo{author}{Li, W.}, \bibinfo{author}{Wang, L.}, \bibinfo{author}{Qin,
  S.}, \bibinfo{year}{2020}b.
\newblock \bibinfo{title}{Cms-unet: Cardiac multi-task segmentation in {MRI}
  with a u-shaped network}, in: \bibinfo{booktitle}{Myocardial Pathology
  Segmentation Combining Multi-Sequence CMR Challenge},
  \bibinfo{organization}{Springer}. pp. \bibinfo{pages}{92--101}.
\bibitem[{Liu et~al.(2016)Liu, Hu, Nordbeck, Ertl, St{\"o}rk and
  Weidemann}]{journal/European/liu2016}
\bibinfo{author}{Liu, D.}, \bibinfo{author}{Hu, K.}, \bibinfo{author}{Nordbeck,
  P.}, \bibinfo{author}{Ertl, G.}, \bibinfo{author}{St{\"o}rk, S.},
  \bibinfo{author}{Weidemann, F.}, \bibinfo{year}{2016}.
\newblock \bibinfo{title}{Longitudinal strain bull’s eye plot patterns in
  patients with cardiomyopathy and concentric left ventricular hypertrophy}.
\newblock \bibinfo{journal}{European Journal of Medical Research}
  \bibinfo{volume}{21}, \bibinfo{pages}{1--12}.
\bibitem[{Liu et~al.(2020)Liu, Zhang, Zhan, Gu and Liu}]{conf/MyoPS/liu2020}
\bibinfo{author}{Liu, Y.}, \bibinfo{author}{Zhang, M.}, \bibinfo{author}{Zhan,
  Q.}, \bibinfo{author}{Gu, D.}, \bibinfo{author}{Liu, G.},
  \bibinfo{year}{2020}.
\newblock \bibinfo{title}{Two-stage method for segmentation of the myocardial
  scars and edema on multi-sequence cardiac magnetic resonance}, in:
  \bibinfo{booktitle}{Myocardial Pathology Segmentation Combining
  Multi-Sequence CMR Challenge}, \bibinfo{organization}{Springer}. pp.
  \bibinfo{pages}{26--36}.
\bibitem[{Lu et~al.(2012)Lu, Yang, Connelly, Wright and
  Radau}]{journal/QIMS/lu2012}
\bibinfo{author}{Lu, Y.}, \bibinfo{author}{Yang, Y.},
  \bibinfo{author}{Connelly, K.A.}, \bibinfo{author}{Wright, G.A.},
  \bibinfo{author}{Radau, P.E.}, \bibinfo{year}{2012}.
\newblock \bibinfo{title}{Automated quantification of myocardial infarction
  using graph cuts on contrast delayed enhanced magnetic resonance images}.
\newblock \bibinfo{journal}{Quantitative imaging in medicine and surgery}
  \bibinfo{volume}{2}, \bibinfo{pages}{81}.
\bibitem[{Ma(2020)}]{conf/MyoPS/ma2020}
\bibinfo{author}{Ma, J.}, \bibinfo{year}{2020}.
\newblock \bibinfo{title}{Cascaded framework with complementary {CMR}
  information for myocardial pathology segmentation}, in:
  \bibinfo{booktitle}{Myocardial Pathology Segmentation Combining
  Multi-Sequence CMR Challenge}, \bibinfo{organization}{Springer}. pp.
  \bibinfo{pages}{159--166}.
\bibitem[{Maier et~al.(2017)Maier, Menze, von~der Gablentz, H{\"a}ni, Heinrich,
  Liebrand, Winzeck, Basit, Bentley, Chen et~al.}]{journal/MedIA/maier2017}
\bibinfo{author}{Maier, O.}, \bibinfo{author}{Menze, B.H.},
  \bibinfo{author}{von~der Gablentz, J.}, \bibinfo{author}{H{\"a}ni, L.},
  \bibinfo{author}{Heinrich, M.P.}, \bibinfo{author}{Liebrand, M.},
  \bibinfo{author}{Winzeck, S.}, \bibinfo{author}{Basit, A.},
  \bibinfo{author}{Bentley, P.}, \bibinfo{author}{Chen, L.}, et~al.,
  \bibinfo{year}{2017}.
\newblock \bibinfo{title}{{ISLES} 2015-a public evaluation benchmark for
  ischemic stroke lesion segmentation from multispectral {MRI}}.
\newblock \bibinfo{journal}{Medical Image Analysis} \bibinfo{volume}{35},
  \bibinfo{pages}{250--269}.
\bibitem[{Mart{\'\i}n-Isla et~al.(2020)Mart{\'\i}n-Isla, Asadi-Aghbolaghi,
  Gkontra, Campello, Escalera and Lekadir}]{conf/MyoPS/martin2020}
\bibinfo{author}{Mart{\'\i}n-Isla, C.}, \bibinfo{author}{Asadi-Aghbolaghi, M.},
  \bibinfo{author}{Gkontra, P.}, \bibinfo{author}{Campello, V.M.},
  \bibinfo{author}{Escalera, S.}, \bibinfo{author}{Lekadir, K.},
  \bibinfo{year}{2020}.
\newblock \bibinfo{title}{Stacked {BCDU-N}et with semantic {CMR} synthesis:
  Application to myocardial pathology segmentation challenge}, in:
  \bibinfo{booktitle}{Myocardial Pathology Segmentation Combining
  Multi-Sequence CMR Challenge}, \bibinfo{organization}{Springer}. pp.
  \bibinfo{pages}{1--16}.
\bibitem[{Menze et~al.(2014)Menze, Jakab, Bauer, Kalpathy-Cramer, Farahani,
  Kirby, Burren, Porz, Slotboom, Wiest et~al.}]{journal/TMI/menze2014}
\bibinfo{author}{Menze, B.H.}, \bibinfo{author}{Jakab, A.},
  \bibinfo{author}{Bauer, S.}, \bibinfo{author}{Kalpathy-Cramer, J.},
  \bibinfo{author}{Farahani, K.}, \bibinfo{author}{Kirby, J.},
  \bibinfo{author}{Burren, Y.}, \bibinfo{author}{Porz, N.},
  \bibinfo{author}{Slotboom, J.}, \bibinfo{author}{Wiest, R.}, et~al.,
  \bibinfo{year}{2014}.
\newblock \bibinfo{title}{The multimodal brain tumor image segmentation
  benchmark ({BRATS})}.
\newblock \bibinfo{journal}{IEEE Transactions on Medical Imaging}
  \bibinfo{volume}{34}, \bibinfo{pages}{1993--2024}.
\bibitem[{Moccia et~al.(2019)Moccia, Banali, Martini, Muscogiuri, Pontone, Pepi
  and Caiani}]{journal/MRMPBM/moccia2019}
\bibinfo{author}{Moccia, S.}, \bibinfo{author}{Banali, R.},
  \bibinfo{author}{Martini, C.}, \bibinfo{author}{Muscogiuri, G.},
  \bibinfo{author}{Pontone, G.}, \bibinfo{author}{Pepi, M.},
  \bibinfo{author}{Caiani, E.G.}, \bibinfo{year}{2019}.
\newblock \bibinfo{title}{Development and testing of a deep learning-based
  strategy for scar segmentation on {CMR-LGE} images}.
\newblock \bibinfo{journal}{Magnetic Resonance Materials in Physics, Biology
  and Medicine} \bibinfo{volume}{32}, \bibinfo{pages}{187--195}.
\bibitem[{Moghari et~al.(2016)Moghari, Pace, Akhondi-Asl and
  Powell}]{link/HVSMR2016}
\bibinfo{author}{Moghari, M.H.}, \bibinfo{author}{Pace, D.F.},
  \bibinfo{author}{Akhondi-Asl, A.}, \bibinfo{author}{Powell, A.J.},
  \bibinfo{year}{2016}.
\newblock \bibinfo{title}{{HVSMR} 2016: {MICCAI} workshop on whole-heart and
  great vessel segmentation from 3{D} cardiovascular {MRI} in congenital heart
  disease}.
\newblock \bibinfo{howpublished}{\url{http://segchd.csail.mit.edu/index.html}}.
\bibitem[{Park et~al.(2019)Park, Liu, Wang and Zhu}]{conf/CVPR/park2019}
\bibinfo{author}{Park, T.}, \bibinfo{author}{Liu, M.Y.}, \bibinfo{author}{Wang,
  T.C.}, \bibinfo{author}{Zhu, J.Y.}, \bibinfo{year}{2019}.
\newblock \bibinfo{title}{Semantic image synthesis with spatially-adaptive
  normalization}, in: \bibinfo{booktitle}{Proceedings of the IEEE/CVF
  Conference on Computer Vision and Pattern Recognition}, pp.
  \bibinfo{pages}{2337--2346}.
\bibitem[{Petitjean et~al.(2015)Petitjean, Zuluaga, Bai, Dacher, Grosgeorge,
  Caudron, Ruan, Ayed, Cardoso, Chen et~al.}]{journal/MedIA/Petitjean2015}
\bibinfo{author}{Petitjean, C.}, \bibinfo{author}{Zuluaga, M.A.},
  \bibinfo{author}{Bai, W.}, \bibinfo{author}{Dacher, J.N.},
  \bibinfo{author}{Grosgeorge, D.}, \bibinfo{author}{Caudron, J.},
  \bibinfo{author}{Ruan, S.}, \bibinfo{author}{Ayed, I.B.},
  \bibinfo{author}{Cardoso, M.J.}, \bibinfo{author}{Chen, H.C.}, et~al.,
  \bibinfo{year}{2015}.
\newblock \bibinfo{title}{Right ventricle segmentation from cardiac {MRI}: a
  collation study}.
\newblock \bibinfo{journal}{Medical Image Analysis} \bibinfo{volume}{19},
  \bibinfo{pages}{187--202}.
\bibitem[{Pizer et~al.(1987)Pizer, Amburn, Austin, Cromartie, Geselowitz,
  Greer, ter Haar~Romeny, Zimmerman and Zuiderveld}]{conf/CVGIP/pizer1987}
\bibinfo{author}{Pizer, S.M.}, \bibinfo{author}{Amburn, E.P.},
  \bibinfo{author}{Austin, J.D.}, \bibinfo{author}{Cromartie, R.},
  \bibinfo{author}{Geselowitz, A.}, \bibinfo{author}{Greer, T.},
  \bibinfo{author}{ter Haar~Romeny, B.}, \bibinfo{author}{Zimmerman, J.B.},
  \bibinfo{author}{Zuiderveld, K.}, \bibinfo{year}{1987}.
\newblock \bibinfo{title}{Adaptive histogram equalization and its variations}.
\newblock \bibinfo{journal}{Computer vision, graphics, and image processing}
  \bibinfo{volume}{39}, \bibinfo{pages}{355--368}.
\bibitem[{Radau et~al.(2009)Radau, Lu, Connelly, Paul, Dick and
  Wright}]{journal/ij/radau2009}
\bibinfo{author}{Radau, P.}, \bibinfo{author}{Lu, Y.},
  \bibinfo{author}{Connelly, K.}, \bibinfo{author}{Paul, G.},
  \bibinfo{author}{Dick, A.}, \bibinfo{author}{Wright, G.},
  \bibinfo{year}{2009}.
\newblock \bibinfo{title}{Evaluation framework for algorithms segmenting short
  axis cardiac {MRI}}.
\newblock \bibinfo{journal}{The MIDAS Journal-Cardiac MR Left Ventricle
  Segmentation Challenge} \bibinfo{volume}{49}.
\bibitem[{Ruder et~al.(2013)Ruder, Ebert, Khattab, Rieben, Thali and
  Kamat}]{journal/FCMP/ruder2013}
\bibinfo{author}{Ruder, T.D.}, \bibinfo{author}{Ebert, L.C.},
  \bibinfo{author}{Khattab, A.A.}, \bibinfo{author}{Rieben, R.},
  \bibinfo{author}{Thali, M.J.}, \bibinfo{author}{Kamat, P.},
  \bibinfo{year}{2013}.
\newblock \bibinfo{title}{Edema is a sign of early acute myocardial infarction
  on post-mortem magnetic resonance imaging}.
\newblock \bibinfo{journal}{Forensic Science, Medicine, and Pathology}
  \bibinfo{volume}{9}, \bibinfo{pages}{501--505}.
\bibitem[{Sandfort et~al.(2017)Sandfort, Kwan, Elumogo, Vigneault, Symons,
  Pourmorteza, Rice, Davies-Venn, Ahlman, Liu
  et~al.}]{journal/JCCT/sandfort2017}
\bibinfo{author}{Sandfort, V.}, \bibinfo{author}{Kwan, A.C.},
  \bibinfo{author}{Elumogo, C.}, \bibinfo{author}{Vigneault, D.M.},
  \bibinfo{author}{Symons, R.}, \bibinfo{author}{Pourmorteza, A.},
  \bibinfo{author}{Rice, K.}, \bibinfo{author}{Davies-Venn, C.},
  \bibinfo{author}{Ahlman, M.A.}, \bibinfo{author}{Liu, C.Y.}, et~al.,
  \bibinfo{year}{2017}.
\newblock \bibinfo{title}{Automatic high-resolution infarct detection using
  volumetric multiphase dual-energy {CT}}.
\newblock \bibinfo{journal}{Journal of cardiovascular computed tomography}
  \bibinfo{volume}{11}, \bibinfo{pages}{288--294}.
\bibitem[{Sedgwick(2014)}]{journal/BMJ/sedgwick2014}
\bibinfo{author}{Sedgwick, P.}, \bibinfo{year}{2014}.
\newblock \bibinfo{title}{Spearman’s rank correlation coefficient}.
\newblock \bibinfo{journal}{Bmj} \bibinfo{volume}{349}.
\bibitem[{Suinesiaputra et~al.(2011)Suinesiaputra, Cowan, Finn, Fonseca,
  Kadish, Lee, Medrano-Gracia, Warfield, Tao and
  Young}]{conf/stacom/Suinesiaputra2011}
\bibinfo{author}{Suinesiaputra, A.}, \bibinfo{author}{Cowan, B.R.},
  \bibinfo{author}{Finn, J.P.}, \bibinfo{author}{Fonseca, C.G.},
  \bibinfo{author}{Kadish, A.H.}, \bibinfo{author}{Lee, D.C.},
  \bibinfo{author}{Medrano-Gracia, P.}, \bibinfo{author}{Warfield, S.K.},
  \bibinfo{author}{Tao, W.}, \bibinfo{author}{Young, A.A.},
  \bibinfo{year}{2011}.
\newblock \bibinfo{title}{Left ventricular segmentation challenge from cardiac
  {MRI}: a collation study}, in: \bibinfo{booktitle}{International Workshop on
  Statistical Atlases and Computational Models of the Heart}, pp.
  \bibinfo{pages}{88--97}.
\bibitem[{Takahashi et~al.(2019)Takahashi, Matsubara and
  Uehara}]{conf/TCSVT/takahashi2019}
\bibinfo{author}{Takahashi, R.}, \bibinfo{author}{Matsubara, T.},
  \bibinfo{author}{Uehara, K.}, \bibinfo{year}{2019}.
\newblock \bibinfo{title}{Data augmentation using random image cropping and
  patching for deep cnns}.
\newblock \bibinfo{journal}{IEEE Transactions on Circuits and Systems for Video
  Technology} \bibinfo{volume}{30}, \bibinfo{pages}{2917--2931}.
\bibitem[{Tan and Le(2019)}]{conf/ICML/tan2019}
\bibinfo{author}{Tan, M.}, \bibinfo{author}{Le, Q.}, \bibinfo{year}{2019}.
\newblock \bibinfo{title}{Efficientnet: Rethinking model scaling for
  convolutional neural networks}, in: \bibinfo{booktitle}{International
  Conference on Machine Learning}, \bibinfo{organization}{PMLR}. pp.
  \bibinfo{pages}{6105--6114}.
\bibitem[{Tao et~al.(2010)Tao, Milles, Zeppenfeld, Lamb, Bax, Reiber and
  van~der Geest}]{journal/MRM/tao2010}
\bibinfo{author}{Tao, Q.}, \bibinfo{author}{Milles, J.},
  \bibinfo{author}{Zeppenfeld, K.}, \bibinfo{author}{Lamb, H.J.},
  \bibinfo{author}{Bax, J.J.}, \bibinfo{author}{Reiber, J.H.},
  \bibinfo{author}{van~der Geest, R.J.}, \bibinfo{year}{2010}.
\newblock \bibinfo{title}{Automated segmentation of myocardial scar in late
  enhancement {MRI} using combined intensity and spatial information}.
\newblock \bibinfo{journal}{Magnetic Resonance in Medicine}
  \bibinfo{volume}{64}, \bibinfo{pages}{586--594}.
\bibitem[{Thygesen et~al.(2008)Thygesen, Alpert and
  White}]{journal/EHJ/Thygesen2008}
\bibinfo{author}{Thygesen, K.}, \bibinfo{author}{Alpert, J.S.},
  \bibinfo{author}{White, H.D.}, \bibinfo{year}{2008}.
\newblock \bibinfo{title}{Universal definition of myocardial infarction}.
\newblock \bibinfo{journal}{European Heart Journal} \bibinfo{volume}{29},
  \bibinfo{pages}{1209}.
\bibitem[{Tobon-Gomez et~al.(2015)Tobon-Gomez, Geers, Peters, Weese, Pinto,
  Karim, Ammar, Daoudi, Margeta, Sandoval et~al.}]{journal/tmi/Tobon2015}
\bibinfo{author}{Tobon-Gomez, C.}, \bibinfo{author}{Geers, A.J.},
  \bibinfo{author}{Peters, J.}, \bibinfo{author}{Weese, J.},
  \bibinfo{author}{Pinto, K.}, \bibinfo{author}{Karim, R.},
  \bibinfo{author}{Ammar, M.}, \bibinfo{author}{Daoudi, A.},
  \bibinfo{author}{Margeta, J.}, \bibinfo{author}{Sandoval, Z.}, et~al.,
  \bibinfo{year}{2015}.
\newblock \bibinfo{title}{Benchmark for algorithms segmenting the left atrium
  from 3{D} {CT} and {MRI} datasets}.
\newblock \bibinfo{journal}{IEEE Transactions on Medical Imaging}
  \bibinfo{volume}{34}, \bibinfo{pages}{1460--1473}.
\bibitem[{Vall and Lemaitre(2016)}]{link/I2CVB2016}
\bibinfo{author}{Vall, J.M.}, \bibinfo{author}{Lemaitre, G.},
  \bibinfo{year}{2016}.
\newblock \bibinfo{title}{I2cvb: initiative for collaborative computer vision
  benchmark}.
\newblock \bibinfo{howpublished}{\url{https://i2cvb.github.io/}}.
\bibitem[{Xiong et~al.(2020)Xiong, Xia, Hu, Huang, Bian, Zheng, Vesal,
  Ravikumar, Maier, Yang et~al.}]{journal/MedIA/xiong2020}
\bibinfo{author}{Xiong, Z.}, \bibinfo{author}{Xia, Q.}, \bibinfo{author}{Hu,
  Z.}, \bibinfo{author}{Huang, N.}, \bibinfo{author}{Bian, C.},
  \bibinfo{author}{Zheng, Y.}, \bibinfo{author}{Vesal, S.},
  \bibinfo{author}{Ravikumar, N.}, \bibinfo{author}{Maier, A.},
  \bibinfo{author}{Yang, X.}, et~al., \bibinfo{year}{2020}.
\newblock \bibinfo{title}{A global benchmark of algorithms for segmenting the
  left atrium from late gadolinium-enhanced cardiac magnetic resonance
  imaging}.
\newblock \bibinfo{journal}{Medical Image Analysis} \bibinfo{volume}{67},
  \bibinfo{pages}{101832}.
\bibitem[{Xu et~al.(2018)Xu, Xu, Gao, Zhao, Zhang, Zhang, Du, Zhao, Ghista, Liu
  et~al.}]{journal/MedIA/xu2018}
\bibinfo{author}{Xu, C.}, \bibinfo{author}{Xu, L.}, \bibinfo{author}{Gao, Z.},
  \bibinfo{author}{Zhao, S.}, \bibinfo{author}{Zhang, H.},
  \bibinfo{author}{Zhang, Y.}, \bibinfo{author}{Du, X.}, \bibinfo{author}{Zhao,
  S.}, \bibinfo{author}{Ghista, D.}, \bibinfo{author}{Liu, H.}, et~al.,
  \bibinfo{year}{2018}.
\newblock \bibinfo{title}{Direct delineation of myocardial infarction without
  contrast agents using a joint motion feature learning architecture}.
\newblock \bibinfo{journal}{Medical Image Analysis} \bibinfo{volume}{50},
  \bibinfo{pages}{82--94}.
\bibitem[{Yu et~al.(2020)Yu, Zha, Huangfu, Chen, Ding and
  Li}]{conf/MyoPS/yu2020}
\bibinfo{author}{Yu, H.}, \bibinfo{author}{Zha, S.}, \bibinfo{author}{Huangfu,
  Y.}, \bibinfo{author}{Chen, C.}, \bibinfo{author}{Ding, M.},
  \bibinfo{author}{Li, J.}, \bibinfo{year}{2020}.
\newblock \bibinfo{title}{Dual attention u-net for multi-sequence cardiac {MR}
  images segmentation}, in: \bibinfo{booktitle}{Myocardial Pathology
  Segmentation Combining Multi-Sequence CMR Challenge},
  \bibinfo{organization}{Springer}. pp. \bibinfo{pages}{118--127}.
\bibitem[{Yushkevich et~al.(2006)Yushkevich, Piven, Hazlett, Smith, Ho, Gee and
  Gerig}]{journal/Neuroimage/Yushkevich2006}
\bibinfo{author}{Yushkevich, P.A.}, \bibinfo{author}{Piven, J.},
  \bibinfo{author}{Hazlett, H.C.}, \bibinfo{author}{Smith, R.G.},
  \bibinfo{author}{Ho, S.}, \bibinfo{author}{Gee, J.C.},
  \bibinfo{author}{Gerig, G.}, \bibinfo{year}{2006}.
\newblock \bibinfo{title}{User-guided 3{D} active contour segmentation of
  anatomical structures: significantly improved efficiency and reliability}.
\newblock \bibinfo{journal}{Neuroimage} \bibinfo{volume}{31},
  \bibinfo{pages}{1116--1128}.
\bibitem[{Zabihollahy et~al.(2019)Zabihollahy, White and
  Ukwatta}]{journal/MP/zabihollahy2019}
\bibinfo{author}{Zabihollahy, F.}, \bibinfo{author}{White, J.A.},
  \bibinfo{author}{Ukwatta, E.}, \bibinfo{year}{2019}.
\newblock \bibinfo{title}{Convolutional neural network-based approach for
  segmentation of left ventricle myocardial scar from 3d late gadolinium
  enhancement {MR} images}.
\newblock \bibinfo{journal}{Medical physics} \bibinfo{volume}{46},
  \bibinfo{pages}{1740--1751}.
\bibitem[{Zhai et~al.(2020)Zhai, Gu, Lei and Wang}]{conf/MyoPS/zhai2020}
\bibinfo{author}{Zhai, S.}, \bibinfo{author}{Gu, R.}, \bibinfo{author}{Lei,
  W.}, \bibinfo{author}{Wang, G.}, \bibinfo{year}{2020}.
\newblock \bibinfo{title}{Myocardial edema and scar segmentation using a
  coarse-to-fine framework with weighted ensemble}, in:
  \bibinfo{booktitle}{Myocardial Pathology Segmentation Combining
  Multi-Sequence CMR Challenge}, \bibinfo{organization}{Springer}. pp.
  \bibinfo{pages}{49--59}.
\bibitem[{Zhang et~al.(2020a)Zhang, Xie, Liao, Verjans and
  Xia}]{conf/MyoPS/zhang2020}
\bibinfo{author}{Zhang, J.}, \bibinfo{author}{Xie, Y.}, \bibinfo{author}{Liao,
  Z.}, \bibinfo{author}{Verjans, J.}, \bibinfo{author}{Xia, Y.},
  \bibinfo{year}{2020}a.
\newblock \bibinfo{title}{Efficientseg: A simple but efficient solution to
  myocardial pathology segmentation challenge}, in:
  \bibinfo{booktitle}{Myocardial Pathology Segmentation Combining
  Multi-Sequence CMR Challenge}, \bibinfo{organization}{Springer}. pp.
  \bibinfo{pages}{17--25}.
\bibitem[{Zhang et~al.(2020b)Zhang, Noga and
  Punithakumar}]{conf/MyoPS/Kumaradevan2020}
\bibinfo{author}{Zhang, X.}, \bibinfo{author}{Noga, M.},
  \bibinfo{author}{Punithakumar, K.}, \bibinfo{year}{2020}b.
\newblock \bibinfo{title}{Fully automated deep learning based segmentation of
  normal, infarcted and edema regions from multiple cardiac {MRI} sequences},
  in: \bibinfo{booktitle}{Myocardial Pathology Segmentation Combining
  Multi-Sequence CMR Challenge}, \bibinfo{organization}{Springer}. pp.
  \bibinfo{pages}{82--91}.
\bibitem[{Zhang et~al.(2020c)Zhang, Liu, Ding, Wang, Pei, Yang and
  Huang}]{conf/MyoPS/zhangzhen2020}
\bibinfo{author}{Zhang, Z.}, \bibinfo{author}{Liu, C.}, \bibinfo{author}{Ding,
  W.}, \bibinfo{author}{Wang, S.}, \bibinfo{author}{Pei, C.},
  \bibinfo{author}{Yang, M.}, \bibinfo{author}{Huang, L.},
  \bibinfo{year}{2020}c.
\newblock \bibinfo{title}{Multi-modality pathology segmentation framework:
  Application to cardiac magnetic resonance images}, in:
  \bibinfo{booktitle}{Myocardial Pathology Segmentation Combining
  Multi-Sequence CMR Challenge}, \bibinfo{organization}{Springer}. pp.
  \bibinfo{pages}{37--48}.
\bibitem[{Zhao et~al.(2020)Zhao, Boutry and Puybareau}]{conf/MyoPS/zhao2020}
\bibinfo{author}{Zhao, Z.}, \bibinfo{author}{Boutry, N.},
  \bibinfo{author}{Puybareau, {\'E}.}, \bibinfo{year}{2020}.
\newblock \bibinfo{title}{Stacked and parallel {U}-nets with multi-output for
  myocardial pathology segmentation}, in: \bibinfo{booktitle}{Myocardial
  Pathology Segmentation Combining Multi-Sequence CMR Challenge},
  \bibinfo{organization}{Springer}. pp. \bibinfo{pages}{138--145}.
\bibitem[{Zhuang(2016)}]{conf/miccai/Zhuang16}
\bibinfo{author}{Zhuang, X.}, \bibinfo{year}{2016}.
\newblock \bibinfo{title}{Multivariate mixture model for cardiac segmentation
  from multi-sequence {MRI}}, in: \bibinfo{booktitle}{International Conference
  on Medical Image Computing and Computer-Assisted Intervention},
  \bibinfo{organization}{Springer}. pp. \bibinfo{pages}{581--588}.
\bibitem[{Zhuang(2019)}]{journal/pami/Zhuang2019}
\bibinfo{author}{Zhuang, X.}, \bibinfo{year}{2019}.
\newblock \bibinfo{title}{Multivariate mixture model for myocardial
  segmentation combining multi-source images}.
\newblock \bibinfo{journal}{IEEE Transactions on Pattern Analysis and Machine
  Intelligence} \bibinfo{volume}{41}, \bibinfo{pages}{2933 -- 2946}.
\bibitem[{Zhuang et~al.(2019)Zhuang, Li, Payer, {\v{S}}tern, Urschler,
  Heinrich, Oster, Wang, Smedby, Bian et~al.}]{journal/MedIA/zhuang2019}
\bibinfo{author}{Zhuang, X.}, \bibinfo{author}{Li, L.}, \bibinfo{author}{Payer,
  C.}, \bibinfo{author}{{\v{S}}tern, D.}, \bibinfo{author}{Urschler, M.},
  \bibinfo{author}{Heinrich, M.P.}, \bibinfo{author}{Oster, J.},
  \bibinfo{author}{Wang, C.}, \bibinfo{author}{Smedby, {\"O}.},
  \bibinfo{author}{Bian, C.}, et~al., \bibinfo{year}{2019}.
\newblock \bibinfo{title}{Evaluation of algorithms for multi-modality whole
  heart segmentation: An open-access grand challenge}.
\newblock \bibinfo{journal}{Medical Image Analysis} \bibinfo{volume}{58},
  \bibinfo{pages}{101537}.
\bibitem[{Zhuang et~al.(2011)Zhuang, Shi, Duckett, Wang, Razavi, Hawkes,
  Rueckert and Ourselin}]{conf/FIMH/zhuang2011}
\bibinfo{author}{Zhuang, X.}, \bibinfo{author}{Shi, W.},
  \bibinfo{author}{Duckett, S.}, \bibinfo{author}{Wang, H.},
  \bibinfo{author}{Razavi, R.}, \bibinfo{author}{Hawkes, D.},
  \bibinfo{author}{Rueckert, D.}, \bibinfo{author}{Ourselin, S.},
  \bibinfo{year}{2011}.
\newblock \bibinfo{title}{A framework combining multi-sequence {MRI} for fully
  automated quantitative analysis of cardiac global and regional functions},
  in: \bibinfo{booktitle}{International Conference on Functional Imaging and
  Modeling of the Heart}, \bibinfo{organization}{Springer}. pp.
  \bibinfo{pages}{367--374}.
\bibitem[{Zhuang et~al.(2020)Zhuang, Xu, Luo, Chen, Ouyang, Rueckert, Campello,
  Lekadir, Vesal, RaviKumar et~al.}]{journal/MedIA/zhuang2020}
\bibinfo{author}{Zhuang, X.}, \bibinfo{author}{Xu, J.}, \bibinfo{author}{Luo,
  X.}, \bibinfo{author}{Chen, C.}, \bibinfo{author}{Ouyang, C.},
  \bibinfo{author}{Rueckert, D.}, \bibinfo{author}{Campello, V.M.},
  \bibinfo{author}{Lekadir, K.}, \bibinfo{author}{Vesal, S.},
  \bibinfo{author}{RaviKumar, N.}, et~al., \bibinfo{year}{2020}.
\newblock \bibinfo{title}{Cardiac segmentation on late gadolinium enhancement
  {MRI}: a benchmark study from multi-sequence cardiac {MR} segmentation
  challenge}.
\newblock \bibinfo{journal}{arXiv preprint arXiv:2006.12434} .

\end{thebibliography}

\end{document}